\documentclass[11pt,oneside]{article}
\usepackage{a4wide}
\usepackage{mathrsfs}
\usepackage{latexsym,bm}
\usepackage{graphicx}
\usepackage{indentfirst}
\usepackage{amsmath}
\usepackage{amssymb}
\usepackage{color}
\usepackage{hyperref}
\usepackage{epsfig}
\usepackage[titletoc]{appendix}
\usepackage{multirow}%
\usepackage{rotating}
\usepackage{epstopdf}
\usepackage{cite}
\usepackage{extarrows}

\newcommand{\be}{\begin{equation}} \newcommand{\ee}{\end{equation}}
\newcommand{\ba}{\begin{array}{c}} \newcommand{\ea}{\end{array}}
\newcommand{\bea}{\begin{eqnarray}} \newcommand{\eea}{\end{eqnarray}}

\newcommand{\order}[1]{\mathcal{O}\left(#1\right)}
\newcommand{\al}{&\!\!\!\!}

\newcommand{\ds}{D_{s0}^*(2317)}
\newcommand{\tadpole}{\mathcal{I}}

\newcommand{\email}[1]{\footnote{{\em E-mail address:} \texttt{#1}}}

\begin{document}
\thispagestyle{empty}

\title{
\LARGE \bf
One-loop analysis of the interactions between charmed mesons
and Goldstone bosons }
\author{De-Liang Yao$^{a,}$\email{d.yao@fz-juelich.de},
Meng-Lin Du$^{b,}$\email{du@hiskp.uni-bonn.de},
Feng-Kun Guo$^{c,b,}\email{fkguo@itp.ac.cn}$,
Ulf-G.~Mei{\ss}ner$^{b,a,}$\email{meissner@hiskp.uni-bonn.de}\\[2mm]
{\small\it  $^a$Institute for Advanced Simulation, Institut f{\"u}r
Kernphysik and J\"ulich Center for Hadron Physics,}\\
{\small\it Forschungszentrum
J{\"u}lich, Wilhelm-Johnen-Stra{\ss}e, D-52425 J{\"u}lich, Germany}
\\
{\small\it  $^b$Helmholtz-Institut f\"ur Strahlen- und Kernphysik and
Bethe Center for Theoretical Physics,}\\
{\small\it Universit\"at Bonn, Nu{\ss}allee 14-16, D--53115
Bonn, Germany}
\\
{\small\it  $^c$State Key Laboratory of Theoretical Physics, Institute of Theoretical Physics,}\\
{\small\it  Chinese Academy of Science, Zhong Guan Cun East Street 55, Beijing 100190, China}}

\maketitle

\begin{abstract}

We derive the scattering amplitude for Goldstone bosons of chiral symmetry off
the pseudoscalar charmed mesons up to  leading one-loop order in a covariant
chiral effective field theory, using the so-called extended-on-mass-shell
renormalization scheme.
Then we use unitarized chiral perturbation theory to fit to the available
lattice data of the $S$-wave scattering lengths. The lattice data are well
described. However, most of the low-energy constants determined from the fit
bear large uncertainties. Lattice simulations in more channels are necessary to
pin down these values which can then be used to make predictions in other
processes related by chiral and heavy quark symmetries.

\end{abstract}
{\hspace{1cm}\small PACS numbers: 12.39.Fe, 13.75.Lb, 14.40.Lb}

\newpage

\section{Introduction}

The hadronic interaction between charmed $D$ mesons and the Goldstone bosons
$\phi$ of the spontaneous breaking of chiral symmetry of the strong interaction
($D$-$\phi$ interaction for short hereafter) is important for the understanding
of the chiral dynamics of quantum chromodynamics (QCD) and the interpretation of the
hadron spectrum in the heavy hadron sector.  Many investigations have been
devoted to study it in the last decade, partly triggered by the observation of
the charm-strange meson $D_{s0}^\ast(2317)$ with $J^P=0^+$ in
2003~\cite{Aubert:2003fg,Krokovny:2003zq}.
The $\ds$ couples to the $DK$ channel, and being below the $DK$ threshold it
decays into the isospin breaking channel $D_s\pi$. In order to unravel  its
nature, theorists study the $D$-$\phi$ interaction and intend to extract  the
information encoded in it. For instance, the $D_{s0}^\ast(2317)$ is interpreted
as a $DK$ molecule~\cite{Barnes:2003dj} by using a chiral unitary approach to
the $S$-wave $D$-$\phi$
interaction~\cite{Kolomeitsev:2003ac,Guo:2006fu,Gamermann:2006nm}.
In these works, the leading order (LO) amplitudes from the heavy meson chiral
perturbation theory (ChPT)~\cite{Burdman:1992gh,Wise:1992hn,Yan:1992gz} are used
as the kernels of resummed amplitudes.
Extensions to the next-to-leading order (NLO) can be found in
Refs.~\cite{Hofmann:2003je,Guo:2008gp,Guo:2009ct,Cleven:2010aw,Guo:2015dha}.

Recently, renewed interest was stimulated due to the occurrence of lattice QCD
calculations of the scattering lengths given in
Refs.~\cite{Liu:2008rza,Liu:2012zya}.
In these works, only channels free of disconnected Wick
contractions are calculated, which are $D\pi$ with isospin $I=3/2$, $D\bar K$
with $I=0$ and 1, $D_sK$ and $D_s\pi$.
There have been lattice results on channels with disconnected Wick
contractions, such as $D\pi$ with $I=1/2$~\cite{Mohler:2012na}
and $DK$ with $I=0$~\cite{Mohler:2013rwa}.
With these lattice calculations,  more insights were gained into the nature of
the $\ds$. The $DK$ isoscalar scattering length was calculated indirectly in
Ref.~\cite{Liu:2012zya}, which is consistent with the result from the direct
lattice calculation in Refs.~\cite{Mohler:2013rwa,Lang:2014yfa}.
A reanalysis of the lattice energy levels for the $D^{(*)}K$ lattice
data~\cite{Lang:2014yfa} was performed in Ref.~\cite{Torres:2014vna} in terms of
an auxiliary potential and an extended L\"uscher formula. These results
suggests that the $\ds$ is dominantly a $DK$ hadronic molecule.~\footnote{A method of extracting the probability of a physical state to be a hadronic molecule in lattice using twisted boundary conditions is discussed in Ref.~\cite{Agadjanov:2014ana} where
the $\ds$ is used as an example for illustration. The $D_{s0}^*(2317)$ was also
studied in a finite volume in Ref.~\cite{MartinezTorres:2011pr}.}

The lattice data can be used to determine the low-energy constants (LECs) in the
chiral Lagrangian of higher orders. Especially, the lattice data in
Refs.~\cite{Liu:2008rza,Liu:2012zya} were used in
Refs.~\cite{Guo:2009ct,Liu:2009uz,Geng:2010vw,Wang:2012bu,Liu:2012zya,Altenbuchinger:2013vwa,Altenbuchinger:2013gaa}.
In the majority of those investigations, unitarized extensions of ChPT, see e.g.
Refs.~\cite{Oller:1997ti,Oller:2000fj}, are adopted so that one can consider
larger meson masses and channel couplings. The unitarized chiral perturbation
theory (UChPT) is especially necessary for the chiral extrapolation of
scattering lengths in question since for larger quark (or meson) masses the interaction normally becomes stronger and
could even be nonperturbative. However, in all the calculations in the
framework of unitarized ChPT, the kernel of the resummed amplitude was only
calculated up to NLO at most and is purely tree-level. Here, we will extend the
calculation to the leading one-loop order, which is the next-to-next-to-leading
order (NNLO).

It is well-known that ChPT~\cite{Weinberg:1978kz,Gasser:1983yg,Gasser:1984gg}
has become an useful and standard tool in studying the hadron interaction at low
energies.  Based on Weinberg's power counting rules~\cite{Weinberg:1978kz},
great achievements have been obtained both in the pure mesonic sector and the
one including matter fields such as baryons, the latter known as baryon ChPT.
There is a notable power counting breaking (PCB) issue in baryon
ChPT~\cite{Gasser:1987rb}:
using the dimensional regularization with the modified minimal subtraction
($\overline{\rm MS}$) scheme in calculating  loop integrals, the naive power
counting does not work and all loop diagrams start contributing at
$\order{p^2}$, with $p$ being a small momentum. There have been several
solutions to this problem:
heavy baryon~(HB) approach~\cite{Jenkins:1990jv,Bernard:1992qa}, infrared
regularizaion (IR)~\cite{Becher:1999he} and extended-on-mass-shell (EOMS)
scheme~\cite{Fuchs:2003qc} (for a review and a detailed comparison of these
approaches, see Ref.~\cite{Bernard:2007zu}).

Likewise, ChPT including the heavy $D$ mesons encounters the same PCB problem.
To remedy it, in Ref.~\cite{Liu:2009uz}, the $D$-$\phi$ scattering lengths were
calculated in the framework of nonrelativistic heavy meson
ChPT~\cite{Wise:1992hn,Yan:1992gz,Burdman:1992gh} to the leading one-loop order
in the heavy quark limit.
Nevertheless, as mentioned by the authors and confirmed by
Ref.~\cite{Geng:2010vw}, this nonrelavistic formulation neglects sizable recoil
corrections\footnote{It is, however, known since a long time that in the heavy
baryon approach such recoil corrections can easily be incorporated by using as
the propagator $i/(v\cdot l -l^2/2m)$ instead of simply $i/v\cdot l$
\cite{Bernard:1993ry}}.  The calculations in various unitarized versions of ChPT
in Refs.~\cite{Guo:2009ct,Liu:2012zya,Wang:2012bu,Altenbuchinger:2013vwa} are
performed in a covariant formalism, but only up to NLO as mentioned above.
The first NNLO calculation of the scattering lengths was given by
Ref.~\cite{Geng:2010vw} using the EOMS scheme. However, the calculation in that
work is perturbative while the interactions in certain channels are definitely
nonperturbative. For instance, in the channel with $(S,I)=(1,0)$, where $S$ and
$I$ represent strangeness and isospin, respectively, the existence of the $\ds$
below the $DK$ threshold calls for a nonperturbative treatment of the $DK$
interaction or inclusion of an explicit field for the $\ds$. In addition, all
the NNLO counterterm contributions are neglected in Ref.~\cite{Geng:2010vw} due
to the poorly known LECs. In this paper, we intend to present a detailed
covariant description of the $D$-$\phi$ interaction up to NNLO in the framework
of UChPT, and the EOMS approach which preserves the proper analytic structure of
the amplitudes will be used in renormalization procedure.

First, we will calculate the $D$-$\phi$ scattering amplitude in covariant ChPT
up to the NNLO.  To our knowledge, the $D$-$\phi$ scattering amplitudes
(without vector charmed mesons) shown in the present work are the first
analytical and complete results up to NNLO.\footnote{The analytical
expressions for the amplitudes involving vector charmed mesons, which survive in
the heavy quark limit, are too lengthy to be shown explicitly in the paper and can
be made available upon request from the authors.} The
vector charmed meson contributions, surviving in the heavy quark limit, are also
taken into account numerically to estimate their influences, although it was
shown in Ref.~\cite{Cleven:2010aw} that their contribution to the $S$-wave scattering
is small.  Renormalization will be performed using the EOMS scheme and it will
be shown explicitly that the UV divergences are cancelled properly and PCB terms
are absorbed exactly by the counterterms, which ensures that the
EOMS-renormalized $D$-$\phi$ scattering amplitudes possess the proper analytic,
power counting, and scale-independent properties.

We will fix the values of the LECs by fitting to the available lattice data of
the $S$-wave $D$-$\phi$ scattering lengths. Since the lattice calculations are
performed at large unphysical quark masses, the perturbative expansion to a
certain order may fail to converge. One way to solve this issue is to employ
unitarized amplitudes instead of  the perturbative ones. Many unitarization
methods have been proposed in the past.
In Ref.~\cite{Oller:2000fj}, a unitarization approach is developed and used to
study the $\bar KN$ interaction. The unitarized amplitude can be matched to the
perturbative amplitude order by order. Throughout this paper, we will call this
approach UChPT for convenience. In Refs.~\cite{Truong:1988zp,Dobado:1989qm}, the
inverse amplitude method (IAM) was proposed and adopted to study the $\pi\pi$
and $K\pi$ scattering. For the purpose of comparison, both the two
approaches will be employed.

This paper is organized as follows. In Section~\ref{SecTheo}, the power counting
and and its breaking by heavy meson masses will be explained briefly and the
chiral effective Lagrangian will be given up to NNLO. In Section~\ref{SecAmpls},
details on the computation of the $D$-$\phi$ scattering amplitude using the EOMS
scheme are exhibited. Together with the loop results shown in
Appendix~\ref{secloopampl}, the minimal but complete set of
scattering amplitudes are given explicitly. Section~\ref{SecUni} discusses how
to obtain the partial wave amplitudes with definite strangeness $S$ and isospin $I$
from the physical process amplitudes, and the two unitarization approaches
mentioned above will be introduced. In Section~\ref{SecSC}, the $S$-wave
scattering lengths are calculated and fitted to the available lattice data at a
few values of the pion masses, and the contributions of the vector charmed
mesons will also be discussed. Finally, Section~\ref{SecCon} comprises a
summary and outlook. Some technicalities are relegated to the appendices.

\section{Theoretical framework\label{SecTheo}}

\subsection{Power counting and power counting breaking terms}

We denote the $D$-$\phi$ interaction as $D_1(p_1)\phi_1(p_2)\to
D_2(p_3)\phi_2(p_4)$.  The scattering process is  on-shell,  hence,
$p_1^2=M_{D_1}^2$,  $p_2^2=M_{\phi_1}^2$, $p_3^2=M_{D_2}^2$ and
$p_4^2=M_{\phi_2}^2$, with $M_{D_1}$ ($M_{\phi_1}$) and $M_{D_2}$ ($M_{\phi_2}$)
being the masses of the incoming and outgoing $D$ mesons (Goldstone bosons),
respectively. In addition, the Mandelstam variables are defined as
\bea
s=(p_1+p_2)^2\ ,\quad t=(p_1-p_3)^2\ ,\quad u=(p_1-p_4)^2\ ,
\eea
which satisfy the relation
$s+t+u=M_{D_1}^2+M_{\phi_1}^2+M_{D_2}^2+M_{\phi_2}^2$.
At low energies, one has
\bea
\frac{s-M_{D_1}^2}{\Lambda_\chi^2}\sim\frac{s-M_{D_2}^2}{\Lambda_\chi^2}
\sim\frac{u-M_{D_1}^2}{\Lambda_\chi^2}\sim\frac{u-M_{D_2}^2}{\Lambda_\chi^2}
\sim\frac{M_{\phi_1}}{\Lambda_\chi}\sim\frac{M_{\phi_2}}{\Lambda_\chi}\ll1\ ,
\quad\frac{t}{\Lambda_\chi^2}\ll1,
\eea
where $\Lambda_\chi\sim \{4\pi F_\pi, M_{D_1},M_{D_2}\}$ denotes the high energy
scale, with $F_\pi$ the pion decay constant $F_\pi \simeq 92.2$~MeV. The
above small quantities can be simultaneously adopted as expansion parameters. In
a more conventional notation, one denotes the small parameters by a unique
symbol, say $p$, so that the power counting rules for the basic quantities read
\bea
&&M_{D_1}\sim O(p^0),\quad M_{D_2}\sim O(p^0),\quad M_{\phi_1}\sim O(p^1),
\quad M_{\phi_2}\sim O(p^1),\quad t\sim O(p^2),\nonumber\\
&&s-M_{D_1}^2\sim O(p^1),\quad s-M_{D_2}^2\sim O(p^1),
\quad u-M_{D_1}^2\sim O(p^1),\quad u-M_{D_2}^2\sim O(p^1).\label{afterrule}
\eea
It is worth noting that the the chiral limit masses of the charmed mesons are of
the same order as the corresponding physical masses. Every physical observable
therefore has its own chiral dimension by using the above given power counting
rules.

Furthermore, in ChPT a power counting rule is assigned  for each Feynman graph.
In the present case, the chiral dimension $n$ for a given graph can be evaluated
from
\bea
n=4\,L+\sum_k\,V_k-2I_\phi-I_D\ ,\label{beforerule}
\eea
where $L$, $V_k$, $I_\phi$ and $I_D$ are the numbers of loops, $k^{\rm th}$
order vertices, Goldstone boson propagators and charmed meson propagators,
respectively.

For a specific physical observable, if there exist terms with chiral dimensions
obtained using Eq.~(\ref{afterrule}) lower than that given by
Eq.~(\ref{beforerule}), those terms are called PCB terms. The PCB terms show up
only when there are heavy particles with nonvanishing chiral limit masses in
loops as internal propagators. In our present calculation, since the heavy
charmed mesons are involved in some of one-loop graphs, there will be PCB terms
if we use dimensional regularization with the $\overline{\rm MS}$ scheme. These
terms can be treated in the so-called EOMS scheme, which has a power counting
consistent with Eq.~\eqref{beforerule}, as will be detailed in
Section~\ref{SecAmpls}.

\subsection{Chiral effective Lagrangian}
The pseudoscalar charmed mesons can be collected in a SU(3) triplet,
$D=(D^0,D^+,D_s^+)$, and the light Goldstone bosons are in an octet,
\bea
\phi=\begin{pmatrix}
   \frac{1}{\sqrt{2}}\pi^0 +\frac{1}{\sqrt{6}}\eta  & \pi^+ &K^+  \\
     \pi^- &  -\frac{1}{\sqrt{2}}\pi^0 +\frac{1}{\sqrt{6}}\eta&K^0 \\
     K^-&\bar{K}^0&-\frac{2}{\sqrt{6}}\eta
\end{pmatrix} .
\eea
The chiral effective Lagrangian for $D$-$\phi$ scattering can be decomposed
into $D$ meson--Goldstone boson interacting parts and pure Goldstone bosonic
parts, which has the following form:
\bea
\mathcal{L}_{\rm eff}=\mathcal{L}^{(1)}_{D\phi}+\mathcal{L}^{(2)}_{D\phi}+
\mathcal{L}^{(3)}_{D\phi}+\mathcal{L}_{\phi\phi}^{(2)}+
\mathcal{L}_{\phi\phi}^{(4)}+\ldots\ .
\eea
Here, the numbers in the superscripts stand for the chiral dimensions, and the
ellipsis denotes the higher-order chiral operators which will not be used here.
Besides, the operators with external fields are also dropped (except for the
scalar external field which is used for the light quark mass insertions).

The familiar lowest order chiral Lagrangian for the Goldstone boson sector reads
\bea
\mathcal{L}_{\phi\phi}^{(2)}=\frac{F_0^2}{4} \left\langle \partial_\mu U
(\partial^\mu U)^\dagger \right\rangle + \frac{F_0^2}{4}\left\langle\chi
U^\dagger+ U\chi^\dagger\right\rangle\ ,
\eea
with $U=\exp\left({i\sqrt{2}\phi}/{F_0}\right)$ and  $\chi=2B_0\,{\rm
diag}(m_u,m_d, m_s)$. Here $\langle\ldots\rangle$ denotes the trace in the
light-flavor space, $F_0$ is the pion decay constant in the chiral
limit, and $B_0$ is a constant related to the quark condensate.
We will work in the isospin limit with $m_u=m_d$ and neglect the electromagnetic
contributions.

The $O(p^4)$ pure Goldstone boson Lagrangian $\mathcal{L}_{\phi\phi}^{(4)}$ is
needed for renormalization. Its LECs enter the $D$-$\phi$ amplitudes
merely through the wave renormalization constants and the decay constants of the
Goldstone bosons, which can be found elsewhere, see  e.g.
Ref.~\cite{Gasser:1984gg}. The relevant terms read
\bea
\mathcal{L}_{\phi\phi}^{(4)}= L_4 \left\langle\partial_\mu U(\partial^\mu
U)^\dagger \right\rangle \left\langle\chi U^\dagger+U\chi^\dagger \right\rangle+
L_5 \left\langle\partial_\mu U(\partial^\mu U)^\dagger \left(\chi
U^\dagger+U\chi^\dagger\right) \right\rangle +\ldots\ .
\eea

For the interaction in the $D$ meson--Goldstone boson sector, the LO effective
Lagrangian takes the form
\bea
\mathcal{L}^{(1)}_{D\phi}=\mathcal{D}_\mu D \mathcal{D}^\mu D^\dagger-{M}_0^2D
D^\dagger\ ,
\eea
where  ${M}_0$ is the mass of the $D$ mesons in the chiral limit, and the
covariant derivative acting on the $D$ mesons is defined by
\bea\label{eq:cov}
\mathcal{D}_\mu D=D(\overset{\leftarrow}{\partial_\mu}+\Gamma_\mu^\dagger)\ ,
\qquad \mathcal{D}_\mu D^\dagger=(\partial_\mu+\Gamma_\mu)D^\dagger\ ,
\eea
with the so-called chiral connection
$
\Gamma_\mu=\left(u^\dagger\partial_\mu u+u\partial_\mu
u^\dagger\right )/2.
$ The NLO Lagrangian reads~\cite{Guo:2008gp}~\footnote{As in
Ref.~\cite{Liu:2012zya}, the $h_6$ term in Ref.~\cite{Guo:2008gp} is dropped,
and the $\tilde{\chi}_+=\chi_+-\langle\chi_+\rangle/3$ is replaced by $\chi_+$
which amounts to a redefinition of $h_0$ and $h_1$. The $h_6$ term is redundant,
since $$-\,h_6\,\mathcal{D}_\mu D[u^\mu,u^\nu]\mathcal{D}_\nu
{D}^\dagger=\frac{h_6}{2}\,\left\{D[u^\mu,u^\nu] (\mathcal{D}_\mu\mathcal{D}_\nu
{D}^\dagger)+(\mathcal{D}_\nu\mathcal{D}_\mu D) [u^\mu,u^\nu] {D}^\dagger\right\}+ {\rm
higher~order~terms},$$  where the first term is zero due to the symmetry
property of the Lorentz indices  $\mu\ ,\,\nu$, and the higher order terms are
contained in the higher order Lagrangians.}
\bea
\mathcal{L}^{(2)}_{D\phi} \al=\al D\left(-h_0\langle\chi_+\rangle-h_1{\chi}_+
+ h_2\langle u_\mu u^\mu\rangle-h_3u_\mu u^\mu\right) {D}^\dag \nonumber\\
\al\al + \mathcal{D}_\mu D\left({h_4}\langle u_\mu
u^\nu\rangle-{h_5}\{u^\mu,u^\nu\}\right)\mathcal{D}_\nu {D}^\dag\ ,
\eea
where the building blocks of the chiral effective Lagrangian are given by
\bea
u_\mu = i\left(u^\dagger\partial_\mu u-u\,\partial_\mu u^\dagger\right)\ ,
\quad u=\exp\left(\frac{i\phi}{\sqrt{2}F_0}\right)\ , \quad
\chi_\pm = u^\dagger\chi u^\dagger\pm u\chi u\ .
\eea
Here, the definition for $\chi_-$ is also given as it  is needed for the NNLO
Lagrangian which, following the procedure detailed in Ref.~\cite{Fettes:2000gb},
can be constructed as
\bea
\mathcal{L}^{(3)}_{D\phi} \al=\al
D\biggl[ i\,{g_1}[{\chi}_-,u_\nu] +
{g_2}\left([u_\mu,[\mathcal{D}_\nu,u^\mu]] + [u_\mu,[\mathcal{D}^\mu,u_\nu]]
\right)\biggr]\mathcal{D}^\nu {D}^\dag \nonumber\\
\al\al + g_3
D\,[u_\mu,[\mathcal{D}_\nu,u_\rho]] \mathcal{D}^{\mu\nu\rho} {D}^\dag\ ,
\eea
where the totally symmetrized  product of three covariant derivatives is defined
as
$\mathcal{D}^{\mu\nu\rho}=\{\mathcal{D}_\mu,
\{\mathcal{D}_\nu,\mathcal{D}_\rho\}\}$.

\section{$D$-$\phi$ scattering amplitudes up to NNLO\label{SecAmpls}}

In this section, we exhibit the complete set of independent $D$-$\phi$
scattering amplitudes on the basis of the physical states. They correspond to 10
physical processes as listed in the second column in Table~\ref{TabCoes1}.
All the other amplitudes can be obtained by using either crossing symmetry or
time-reversal invariance. In what follows, we will first calculate the
tree-level amplitude which can be reduced into a common structure but with
different coefficients because of  SU(3) flavor symmetry. Then the loop
amplitudes will be given explicitly. In the end, the renormalization procedure
within the EOMS scheme will be discussed.

\subsection{Tree-level contribution\label{SecTree}}

\begin{figure}[t]
\begin{center}
\epsfig{file=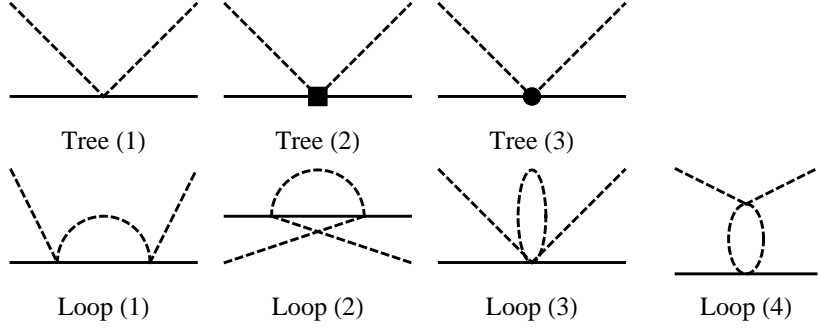,scale=0.55}
\caption{The 1-point irreducible (1PI) Feynman diagrams for $D$-$\phi$
scattering up to leading one-loop order. The solid (dashed) lines represent the
$D$ (Goldstone) mesons. The square stands for the contact vertex coming from
Lagrangian $\mathcal{L}_{D\phi}^{(2)}$, while the filled circle denotes an
insertion from $\mathcal{L}_{D\phi}^{(3)}$. All other vertices are generated
either by $\mathcal{L}_{D\phi}^{(1)}$ or  $\mathcal{L}_{\phi\phi}^{(2)}$.}
\end{center}
\vspace{-5mm}
\label{FeynDiags}
\end{figure}

The Feynman diagrams of the tree-level contribution to the scattering amplitudes
are displayed in the first line of Fig.~\ref{FeynDiags}. Since we do not
consider the exchange of resonances, such contributions are encoded in the
contact terms for the $D$-$\phi$ scattering. When calculating the Feynman
diagrams, all the bare parameters, such as the decay constant $F_0$ and
the masses, are
maintained. They will be replaced by the corresponding physical quantities when
the renormalization is performed. The LO, i.e. $\mathcal{O}(p)$, tree amplitude
is the Weinberg--Tomozawa term~\footnote{As the vector charmed mesons are not
taken into account, there is no Born term due to the exchange of these
mesons.}, and has the following form,
\bea
\mathcal{A}^{(1)}(s,t,u)=\mathcal{C}_\text{LO}\frac{s-u}{4F_0^2}\ ,
\eea
where the coefficients $\mathcal{C}_\text{LO}$ for different physical processes
are listed in Table~\ref{TabCoes1}. The Weinberg-Tomozawa term depends only on
the pion decay constant due to the fact that it originates from the kinetic term
in $\mathcal{L}_{D\phi}^{(1)}$, which is a result of the spontaneous breaking of
chiral symmetry in QCD.

The $\mathcal{O}(p^2)$ Lagrangian $\mathcal{L}_{D\phi}^{(2)}$ generates the
tree-level contribution at NLO as
\bea
\mathcal{A}^{(2)}(s,t,u)=\frac{1}{F_0^2}\left[-4h_0\mathcal{C}_0^{(2)}+
{2}h_1\mathcal{C}_1^{(2)}-2\mathcal{C}_{24}^{(2)}H_{24}(s,t,u)+
2\mathcal{C}_{35}^{(2)}H_{35}(s,t,u)\right]\ ,
\eea
where the coefficients  are shown in Table~\ref{TabCoes1}, and the functions
$H_{24}(s,t,u)$ and $H_{35}(s,t,u)$ are defined by
\bea
H_{24}(s,t,u)\al=\al 2h_2\,p_2\cdot p_4+h_4\,(p_1\cdot p_2 p_3\cdot p_4+p_1\cdot
p_4 p_2\cdot p_3)\ ,\\
H_{35}(s,t,u)\al=\al h_3\,p_2\cdot p_4+h_5\,(p_1\cdot p_2 p_3\cdot p_4+p_1\cdot
p_4 p_2\cdot p_3)\ .
\eea

Finally,  the tree-level amplitude at $\mathcal{O}(p^3)$ reads
\bea
\mathcal{A}^{(3)}(s,t,u) = \frac{1}{F_0^2}
\left\{4g_1 \! \left[\mathcal{C}_{1a}^{(3)}(p_1+p_3)\cdot(p_2+p_4)+
\mathcal{C}_{1b}^{(3)}(p_1+p_3)\cdot p_2\right]\! +
4\mathcal{C}_{23}^{(3)}G_{23}(s,t,u)\right\},
\eea
with
\bea
G_{23}(s,t,u) \al=\al -g_2\,p_2\cdot p_4(p_1+p_3)\cdot(p_2+p_4) \nonumber\\
\al \al + 2g_3\left[ (p_1\cdot p_2)( p_1\cdot p_4) p_1\cdot(p_2+p_4)
+(p_1\to p_3)\right]\ .
\eea
The corresponding coefficients can be found in Table~\ref{TabCoes2}.
The $\mathcal{C}_{1b}^{(3)}$ term survives only for inelastic scattering
processes.

\begin{table}[t]
\caption{The coefficients in the LO and NLO tree-level amplitudes of the 10
relevant physical processes. The Gell-Mann--Okubo mass relation,
$3M_\eta^2=4M_K^2-M_\pi^2$, is used to simplify the coefficients when necessary.
}\label{TabCoes1}
\vspace{-0.5cm}
\bea
\begin{array}{ll|c|cccc}
\hline\hline
&{\rm
Physical~processes}&\mathcal{C}_\text{LO}&\mathcal{C}_{0}^{(2)}&
\mathcal{C}_{1}^{(2)}&\mathcal{C}_{24}^{(2)}&\mathcal{C}_{35}^{(2)}\\
\hline
1&D^0K^-\to D^0K^-& 1 & M_K^2 & -M_K^2 & 1 & 1 \\
2&D^+K^+\to D^+K^+& 0 & M_K^2 & 0 & 1 & 0  \\
3&D^+\pi^+\to D^+\pi^+& 1 & M_{\pi }^2 & -M_{\pi }^2 & 1 & 1\\
4&D^+\eta\to D^+\eta& 0 & M_\eta^2 & -\frac{1}{3}M_{\pi }^2 & 1 & \frac{1}{3} \\
5&D_s^+K^+\to D_s^+K^+& 1 & M_K^2 & -M_K^2 & 1 & 1 \\
6&D_s^+\eta\to D_s^+\eta& 0 &M_\eta^2 &  \frac{4}{3}  \left(M_\pi^2-2 M_{K
}^2\right) & 1 & \frac{4}{3} \\
7&D_s^+\pi^0\to D_s^+\pi^0& 0 & M_{\pi }^2 & 0 & 1 & 0 \\
8&D^0\eta\to D^0\pi^0& 0 & 0 & -\frac{1}{\sqrt{3}} M_{\pi }^2 & 0 & \frac{1}{\sqrt{3}} \\
9&D_s^+K^-\to D^0\pi^0& -\frac{1}{\sqrt{2}} & 0 &  -\frac{1 }{2
\sqrt{2}}\left(M_K^2+M_{\pi }^2\right) & 0 & \frac{1}{\sqrt{2}}  \\
10&D_s^+K^-\to D^0\eta& -\sqrt{\frac{3}{2}} & 0 &  \frac{1}{2\sqrt{6}}
 \left(5 M_K^2-3 M_{\pi }^2\right) & 0 & -\frac{1}{\sqrt{6}}
\\
\hline\hline
\end{array}\nonumber
\eea
\end{table}

\begin{table}[t]
\caption{The coefficients in the NNLO tree-level amplitudes of the 10 relevant
physical processes. }
\label{TabCoes2}
\vspace{-0.5cm}
\bea
\begin{array}{ll|ccc}
\hline\hline
&{\rm
physical~process}&\mathcal{C}_{1a}^{(3)}&\mathcal{C}_{1b}^{(3)}&
\mathcal{C}_{23}^{(3)}\\
\hline
1&D^0K^-\to D^0K^-&  M_K^2 & 0 & 1 \\
2&D^+K^+\to D^+K^+&  0 & 0 & 0 \\
3&D^+\pi^+\to D^+\pi^+&  M_{\pi }^2 & 0 & 1 \\
4&D^+\eta\to D^+\eta&  0 & 0 & 0 \\
5&D_s^+K^+\to D_s^+K^+&  M_K^2 & 0 & 1 \\
6&D_s^+\eta\to D_s^+\eta&  0 & 0 & 0 \\
7&D_s^+\pi^0\to D_s^+\pi^0&  0 & 0 & 0 \\
8&D^0\eta\to D^0\pi^0&  0 & 0 & 0 \\
9&D_s^+K^-\to D^0\pi^0& -\frac{1}{\sqrt{2}}M_K^2 &
\frac{1}{\sqrt{2}}\left(M_K^2-M_{\pi }^2\right) & -\frac{1}{\sqrt{2}} \\
10&D_s^+K^-\to D^0\eta& -\sqrt{\frac{3}{2}} M_K^2 &
\frac{1}{\sqrt{6}}\left(M_{\pi }^2-M_K^2\right) & -\sqrt{\frac{3}{2}} \\
\hline\hline
\end{array}\nonumber
\eea
\end{table}

\subsection{One-loop contribution\label{SecLoop}}

The one-loop connected graphs for $D$-$\phi$ scattering are shown in the second
line of Fig.~\ref{FeynDiags}. All the vertices in the loop graphs originate from
the Lagrangians $\mathcal{L}_{D\phi}^{(1)}$ and $\mathcal{L}_{\phi\phi}^{(2)}$
which are free of unknown LECs. Similar to the tree-level amplitudes, it
suffices to calculate the loop amplitudes for the 10 physical processes. All
these loop amplitudes are listed in  Appendix~\ref{secloopampl}, which are
expressed in terms of a set of one-loop integrals given in
Appendix~\ref{secloopint}.

\subsection{Renormalization}

In the previous sections, the 1PI Feynman graphs are all calculated, which are
related to the so-called amputated amplitudes. To derive the $S$-matrix
elements, one should perform  wave function renormalization. Moreover, in the end, all
the bare parameters should be replaced by the corresponding physical ones.

\subsubsection{Wave function renormalization}

\begin{figure}[htbp]
\begin{center}
\epsfig{file=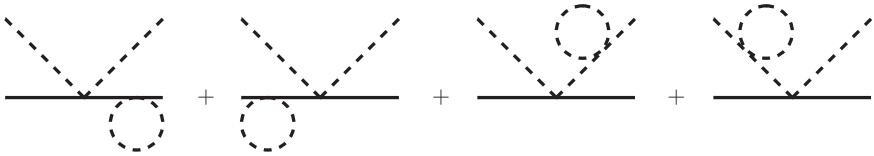,scale=0.618}
\caption{Feynman diagrams for the wave function renormalization at
$\order{p^3}$.}
\label{WaveFun}
\end{center}
\vspace{-5mm}
\end{figure}

To perform the wave function renormalization, one multiplies the external lines
with the square roots of the wave function renormalization constants of the
corresponding fields and takes them on the mass shell. In perturbation theory,
if the calculation is done up to a certain order (up to $\mathcal{O}(p^3)$ in
our case), the wave function renormalization is equivalent to taking the graphs in
Fig.~\ref{WaveFun} into account. All the higher order contributions beyond the
required accuracy are ignored.

Hence, when taking wave function renormalization into consideration, the
scattering amplitude becomes
 \bea
\mathcal{A}(s,t)=\mathcal{A}^{(1)}_{\rm tree}(s,t)+ \mathcal{A}^{(2)}_{\rm
tree}(s,t)+\mathcal{A}^{(3)}_{\rm tree}(s,t)+\mathcal{A}^{(3)}_{\rm loop}(s,t)+
\mathcal{A}^{(3)}_{\rm wf}(s,t)\ . \label{fullamp1}
\eea
The first three terms are tree contribution given in Section~\ref{SecTree},
while the fourth term is the loop contribution discussed in
Section~\ref{SecLoop} and Appendix~\ref{secloopampl}. The last term
$\mathcal{A}_{\rm wf}(s,t)$ corresponds to the contribution from the wave
function renormalization. It can be obtained from the LO amplitude in
combination with the wave function renormalization constants. For instance,
considering the scattering process $D_1\phi_1\to D_2\phi_2$, it is given by
\bea
\mathcal{A}^{(3)}_{\rm wf}(s,t)=\frac{1}{2}(\delta\mathcal{Z}_{D_1}
+\delta\mathcal{Z}_{\phi_1}+\delta\mathcal{Z}_{D_2}
+\delta\mathcal{Z}_{\phi_2})\mathcal{A}^{(1)}_{\rm tree}(s,t)\ ,
\eea
with $\delta\mathcal{Z}=\mathcal{Z}-1$ and $\mathcal{Z}$ being the wave
function renormalization constant up to the order considered. To be explicit,
the wave function renormalization constants for $D$ and $D_s$ are
$\mathcal{Z}_D=\mathcal{Z}_{D_s}=1$ and for the Goldstone bosons are
\bea
\mathcal{Z}_\pi\al=\al
1-\frac{1}{F_0^2}\left[8L_4(2M_K^2+M_\pi^2)+8L_5M_\pi^2+
\frac{1}{3}\tadpole_K+\frac{2}{3}\tadpole_\pi\right] ,\nonumber\\
\mathcal{Z}_K\al=\al
1-\frac{1}{F_0^2}\left[8L_4(2M_K^2+M_\pi^2)+8L_5M_K^2+
\frac{1}{2}\tadpole_K+\frac{1}{4}\tadpole_\pi+\frac{1}{4}\tadpole_\eta\right]
,\nonumber\\
\mathcal{Z}_\eta\al=\al
1-\frac{1}{F_0^2}\left[8L_4(2M_K^2+M_\pi^2)+
\frac{4}{3}L_5(4M_K^2-M_\pi^2)+\tadpole_K\right] ,
\eea
where the tadpole loop integral $\tadpole_i (i=\pi,K,\eta)$ can be found in
Appendix~\ref{secloopint}. Note that in the above expressions, the ultraviolet
(UV) divergence of the loop functions is not subtracted on purpose. This is due
to the fact that the $\mathcal{Z}$'s are not physical observables such that they
might be divergent, namely the LECs $L_4$ and $L_5$ are not
sufficient to absorb the UV divergence in those expressions. The UV divergence
cancellation as well as the PCB terms absorption will be discussed in the
following section at the level of the $S$-matrix elements. As one will see, the
$S$-matrix elements are free of any divergence.

\subsubsection{Extended-On-Mass-Shell subtraction scheme\label{secEOMS}}

The loop integrals in the amplitude shown in Eq.~(\ref{fullamp1}) is UV
divergent, and we need renormalization to absorb the divergences by
counterterms. Moreover, PCB terms show up in the chiral expansion if we use
dimensional regularization with the $\overline{\rm MS}$ scheme. It is necessary
to get rid of them to have a good power counting. We will use the EOMS
subtraction scheme which has the proper analyticity and correct power counting
for the amplitudes. The essence of the EOMS scheme is to perform two subsequent
subtractions: the $\overline{\rm MS}$ subtraction and the EOMS finite
subtraction.

In the $\overline{\rm MS}$ subtraction, the UV divergent parts are extracted and
then cancelled by the divergences in the bare LECs, which are separated into
finite and divergent parts as follows:
\bea
h_i=h_i^r(\mu)+\frac{\alpha_i}{16\pi^2F_0^2}R\ ,\qquad
g_j=g_j^r(\mu)+\frac{\beta_j}{16\pi^2F_0^2}R\ ,\qquad
L_k=L_k^r(\mu)+\frac{\Gamma_k}{32\pi^2}R\ ,\label{MSbarren}
\eea
where  $R=\frac{2}{d-4}+\gamma_E-1-\ln(4\pi)$ with $\gamma_E$ being the Euler
constant, and $d$
is the space-time dimension. The coefficients
$\alpha_{i}$ ($i=0,\cdots,5$), $\beta_{j}$ ($j=1,2,3$) and $\Gamma_{k}$
($k=4,5$) are given by $$\alpha_0=0,\quad \alpha_1=0,\quad
\alpha_2=\frac{M_0^2}{48},\quad \alpha_3=-\frac{M_0^2}{16}
,\quad \alpha_4=\frac{7}{24},\quad \alpha_5=-\frac{7}{16},$$
$$\beta_1=0,\quad \beta_2=-\frac{9}{128},\quad \beta_3=0,\quad
\Gamma_4=\frac{1}{8},\quad \Gamma_5=\frac{3}{8}.$$

Although the UV divergences have been removed so far, it is still not sufficient
to get an amplitude that respects the power counting rule given by
Eq.~(\ref{beforerule}).
The charmed  mesons show up in the Feynman diagrams, say Loop(1) and
Loop(2) in Fig.~\ref{FeynDiags}, and generate the so-called PCB terms that often
spoil the convergence of the chiral expansion~\cite{Gasser:1987rb}. The EOMS
finite subtraction is used to get rid of those PCB terms. For each physical
process given above, the PCB terms are easily obtained by replacing the loop
function $\mathcal{F}_{ab}^{(cd)}(s,t)$, see
Eq.~\eqref{eq:F}, by $\mathcal{F}_{ab}^{(cd)}(s,t)^{\rm PCB}$, namely
Eq.~(\ref{PCBmathF}), in the amplitudes and then performing the chiral expansion
with respect to the small quantities. Note that the infrared regular parts for
the required scalar loop integrals are also listed in Appendix~\ref{secRegular}
for easy reference. Eventually, the PCB terms are absorbed by decomposing the
$\overline{\rm MS}$-renormalized LECs in the $\mathcal{O}(p^2)$ Lagrangian via
\bea
h_i^r(\mu)=\tilde{h}_i+\frac{\delta_i}{16\pi^2F_0^2}M_0^2\ ,\label{EOMSren}
\eea with the coefficient $\delta_i$ ($i=0,\cdots,5$) defined by
\bea
\delta_0=\delta_1=0\ ,\,\delta_2=-\frac{1}{72}+\frac{1}{48}\log\frac{M_0^2}{\mu^2}\ ,\,\delta_3=\frac{1}{24}-\frac{1}{16}\log\frac{M_0^2}{\mu^2}\
,\,\nonumber\\
\delta_4=-\frac{35}{72M_0^2}+\frac{7}{24M_0^2}\log\frac{M_0^2}{\mu^2}\ ,\,\delta_5=\frac{35}{48M_0^2}-\frac{7}{16M_0^2}\log\frac{M_0^2}{\mu^2}\ .
\eea
The other LECs such as $g_j^r(\mu)$ and $L_k^r(\mu)$ are untouched when
performing the finite EOMS subtraction.

After the two steps described above, we have obtained the full
renormalized amplitudes. For the sake of easy practical usage, the
chiral-limit $D$ meson mass $M_0$ and the chiral-limit decay constant $F_0$
should be further related to the corresponding physical quantities according to
the following expressions:
\bea
M_D^2\al=\al {M_0}^2+2\,(h_0+h_1)M_\pi^2+4\,h_0\,M_K^2\ ,\label{Dmass}\\
M_{D_s}^2\al=\al {M_0}^2+2\,(h_0-h_1)M_\pi^2+4\,(h_0+h_1)M_K^2 \
,\label{Dsmass}\\
F_\pi\al=\al
F_0+\frac{1}{2F_0}\left(2\tadpole_\pi^r+\tadpole_K^r\right)+
\frac{4M_\pi^2}{F_0}\left(L_4^r + L_5^r \right)+\frac{8M_K^2}{F_0}L_4^r \ .
\eea
Here, we rewrite $F_0$ in terms of $F_\pi$ rather than $F_K$ and
$F_\eta$. This is the convention to be used throughout.  Alternatively, one
can also rewrites it in terms of $F_K$ or $F_\eta$, and the difference is
of higher order.
The loop functions and LECs with a superscript $r$ stand for their
finite parts, namely, the contributions proportional to the UV divergence $R$
are removed.

\section{Partial wave amplitudes and unitarization\label{SecUni}}

In this section, we will illustrate how to obtain partial wave amplitudes with
definite strangeness $S$ and isospin $I$ from the 10 physical process amplitudes
exhibited in the previous section in detail. Then we will discuss the
unitarization of the scattering amplitudes using two different approaches.
Based on the content of this section, it is straightforward to derive the
$S$-wave scattering lengths, which will be discussed and compared with lattice
data in the next section.

\subsection{Amplitudes for given strangeness and isospin}

The scattering amplitudes in the isospin basis can be classified by two quantum
numbers, which are the strangeness $S$ and isospin $I$ of the scattering system.
Hereafter, the scattering amplitudes with definite strangeness and isospin are
called strangeness-isospin amplitudes for short. All the strangeness-isospin
amplitudes can be related to the 10 amplitudes of the physical processes using
crossing symmetry and isospin symmetry.

We begin with the single-channel interactions. There are 4 single channels in
total. The corresponding quantum numbers of $(S,I)$ are $(-1,0)$, $(-1,1)$,
$(0,3/2)$ and $(2,1/2)$. Their strangeness-isospin amplitudes are related to the
physical-process amplitudes by
\bea
\mathcal{A}^{(-1,0)}_{D\bar{K}\to D\bar{K}}(s,t,u)\al=\al2\,
\mathcal{A}_{D^+K^+\to D^+K^+}(u,t,s)-\mathcal{A}_{D^0K^-\to D^0K^-}(s,t,u)\ ,\\
\mathcal{A}^{(-1,1)}_{D\bar{K}\to D\bar{K}}(s,t,u)\al=\al \mathcal{A}_{D^0K^-\to
D^0K^-}(s,t,u)\ ,\\
\mathcal{A}^{(0,3/2)}_{D\pi\to D\pi}(s,t,u)\al=\al \mathcal{A}_{D^+\pi^+\to
D^+\pi^+}(s,t,u)\ ,\\
\mathcal{A}_{D_sK\to D_s K}^{(2,1/2)}(s,t,u)\al=\al \mathcal{A}_{D_s^+K^+\to
D_s^+K^+}(s,t,u)\ .
\eea
For the coupled channels with $(S,I)=(1,0)$, the strangeness-isospin amplitudes
read
\bea
\mathcal{A}^{(1,0)}_{D{K}\to D{K}}(s,t,u)\al=\al2\, \mathcal{A}_{D^0K^-\to
D^0K^-}(u,t,s)-\mathcal{A}_{D^+K^+\to D^+K^+}(s,t,u)\ ,\\
\mathcal{A}^{(1,0)}_{D_s\eta\to D_s\eta}(s,t,u)\al=\al \mathcal{A}_{D_s^+\eta\to
D_s^+\eta}(s,t,u)\ ,\\
\mathcal{A}^{(1,0)}_{D_s\eta\to DK}(s,t,u)\al=\al
-\sqrt{2}\,\mathcal{A}_{D_s^+K^-\to D^0\eta}(u,t,s)\ .
\eea
For the coupled channels with $(S,I)=(1,1)$, one has
\bea
\mathcal{A}_{D_s\pi\to D_s\pi}^{(1,1)}(s,t,u)\al=\al \mathcal{A}_{D_s^+\pi^0\to
D_s^+\pi^0}(s,t,u)\ ,\\
\mathcal{A}^{(1,1)}_{D{K}\to D{K}}(s,t,u)\al=\al \mathcal{A}_{D^+K^+\to
D^+K^+}(s,t,u)\ ,\\
\mathcal{A}^{(1,1)}_{DK\to D_s\pi}(s,t,u)\al=\al
\sqrt{2}\,\mathcal{A}_{D_s^+K^-\to D^0\pi^0}(u,t,s)\ .
\eea
For $(S,I)=(0,1/2)$, there are three channels: $D\pi,D\eta$ and $D_s\bar K$. The
isospin relations are given by
\bea
\mathcal{A}_{D\pi\to D\pi}^{(0,1/2)}(s,t,u)\al=\al\frac{3}{2}\mathcal{A}_{D^+\pi^+\to D^+\pi^+}(u,t,s)-\frac{1}{2}\mathcal{A}_{D^+\pi^+\to D^+\pi^+}(s,t,u)\ ,\\
\mathcal{A}_{D\eta\to D\eta}^{(0,1/2)}(s,t,u)\al=\al\mathcal{A}_{D^+\eta\to D^+\eta}(s,t,u)\ ,\\
\mathcal{A}_{D_s\bar{K}\to D_s\bar{K}}^{(0,1/2)}(s,t,u)\al=\al\mathcal{A}_{D_s^+K^+\to D_s^+K^+}(u,t,s)\ ,\\
\mathcal{A}_{D\eta\to D\pi}^{(0,1/2)}(s,t,u)\al=\al\sqrt{3}\mathcal{A}_{D^0\eta\to D^0\pi^0}(s,t,u)\ ,\\
\mathcal{A}_{D_s\bar{K}\to D\pi}^{(0,1/2)}(s,t,u)\al=\al\sqrt{3}\mathcal{A}_{D_s^+K^-\to D^0\pi^0}(s,t,u)\ ,\\
\mathcal{A}_{D_s\bar{K}\to D\eta}^{(0,1/2)}(s,t,u)\al=\al\mathcal{A}_{D_s^+K^-\to D^0\eta}(s,t,u)\ .
\eea

\subsection{Partial wave projection}
Each of the strangeness-isospin amplitudes can be denoted by
$\mathcal{A}^{(S,I)}_{D_1\phi_1\to D_2\phi_2}(s,t)$.  Its partial wave
projection with definite angular momentum $\ell$ is given by
\bea
\mathcal{A}_{\ell}^{(S,I)}(s)_{D_1\phi_1\to D_2\phi_2}
= \frac{1}{2}\int_{-1}^1{\rm
d}\cos\theta\,P_\ell(\cos\theta)\, \mathcal{A}^{(S,I)}_{D_1\phi_1\to
D_2\phi_2}(s,t(s,\cos\theta))\ .
\label{eq:pwp}
\eea
Here, the Mandelstam variable $t$ is expressed in terms of $s$ and the
scattering angle $\theta$,
\bea
t(s,\cos\theta)\al=\al M_{D_1}^2+M_{D_2}^2-
\frac{1}{2s}\left(s+M_{D_1}^2-M_{\phi_1}^2\right)
\left(s+M_{D_2}^2-M_{\phi_2}^2\right) \nonumber\\
\al\al
-\frac{\cos\theta}{2s}\sqrt{\lambda(s,M_{D_1}^2,M_{\phi_1}^2)
\lambda(s,M_{D_2}^2,M_{\phi_2}^2)}\ .
\label{eq:t}
\eea
with $\lambda(a,b,c)=a^2+b^2+c^2-2ab-2bc-2ac$  the K\"all\'{e}n function.
From Eq.~\eqref{eq:t}, one sees that at each of the thresholds of $D_1\phi_1$
and $D_2\phi_2$, i.e. when $s$ takes one of the following two values
\bea
s_{1}=(M_{D_1}+M_{\phi_1})^2\ ,\qquad s_2=(M_{D_2}+M_{\phi_2})^2\ ,
\eea
$t$ is independent of $\cos\theta$. Taking $s=s_1$ for instance, the $S$-wave
amplitude becomes
\bea
\mathcal{A}_{\ell=0}^{(S,I)}(s_1)_{D_1\phi_1\to D_2\phi_2}=
\mathcal{A}^{(S,I)}_{D_1\phi_1\to D_2\phi_2}(s_1,t(s_1))\ .\label{Swave}
\eea
This means that the $S$-wave amplitude at threshold can be obtained directly
from the full amplitude by setting the energy squared at its threshold value.
However, note that this simple recipe can only be used for the single
channel case. For coupled channels, it is necessary to perform the partial wave
projection using Eq.~\eqref{eq:pwp}.

Before ending this section, we remark that it is helpful to use matrix notation
to denote the partial wave amplitudes with definite strangeness $S$ and isospin
$I$. In the  matrix notation, the subscript $D_1\phi_1\to D_2\phi_2$ is
redundant. For single channels, this is apparent since the process is specified
uniquely by $(S,I)$. For coupled channels, taking $(S,I)=(1,1)$ for example,
there are four processes: $D_s\pi\to D_s\pi$, $DK\to
DK$, $DK\to D_s\pi$ and its time reversal process. Using time reversal
invariance, one can write
\bea
\mathcal{A}_{\ell}^{(1,1)}(s)=\begin{pmatrix}
      \mathcal{A}_{\ell}^{(1,1)}(s)_{D_s\pi\to D_s\pi}&
      \mathcal{A}_{\ell}^{(1,1)}(s)_{DK\to D_s\pi}    \\
      \mathcal{A}_{\ell}^{(1,1)}(s)_{DK\to D_s\pi} &
     \mathcal{A}_{\ell}^{(1,1)}(s)_{DK\to DK} \end{pmatrix}.
\eea
Later on, we will refer to the amplitudes for a given process in the isospin
basis by $\mathcal{A}_{\ell}^{(S,I)}(s)_{ij}$, with $i$ and $j$ being channel
indices. Unitarization of the scattering amplitudes will be discussed in the
matrix notation in the following.

\subsection{Unitarization}

Unitarization is often adopted to extend ChPT to higher energies.
The unitarized amplitudes sum up a series of $s$-channel
loops,~\footnote{Since the unitarization procedure is normally equivalent to
a resummation of the scattering amplitudes in the $s$-channel, it breaks the
crossing symmetry. Crossing symmetry can be restored using Roy-type equations,
for an early attempt, see
Ref.~\cite{Hannah:2001ee}.} which correspond to the right-hand cut, and thus
one would naively expect that they can be used for higher momenta as well as
larger pion masses. Phenomenologically, it is now well-known that the unitarized
amplitudes can well describe the scattering data for the pion and kaon systems up to
1.2~GeV, see, e.g., Refs.~\cite{Oller:1997ti,Oller:1998hw}. We thus expect that
these amplitudes allows for a description of the lattice data at pion masses
higher than the conventional ChPT. Yet, there is no rigorous proof a priori.
For varying the quark masses (or equivalently the masses of the Goldstone
bosons), it provides a way to performing the chiral extrapolation  of lattice
simulation results or studying the quark mass dependence of physical quantities.
In the present work, we will consider two different versions of unitarization
for the sake of comparison and for quantifying the inherent model-dependence of
such approaches. For the sake of simplicity and generality, all the quantum
number indices of the amplitudes such as $S$, $I$ and $\ell$ will be suppressed
in this section.
That is to say $T$, $\mathcal{A}$, $\mathcal{T}$ and $\tilde{T}$, which will appear later on,  are
$T^{(S,I)}_{\ell}$, $\mathcal{A}^{(S,I)}_{\ell}$, $\mathcal{T}^{(S,I)}_{\ell}$
and $\tilde{T}^{(S,I)}_{\ell}$, respectively, for our case.

The first approach we will use is the one proposed in Ref.~\cite{Oller:2000fj},
which is denoted by UChPT throughout this paper. In matrix form, the unitarized
amplitude is given by
\bea
T(s)=\left\{1-\mathcal{T}(s)\cdot g(s)\right\}^{-1}\cdot\mathcal{T}(s)\ ,
\label{UniUChPT}
\eea
where $g(s)$ is a diagonal matrix $g(s)={\rm diag}\{g(s)_i\}$, with $i$ the
channel index. The fundamental loop integral  $g(s)_i$ reads
\bea
g(s)_i=i\int\frac{{\rm d}^4q}{(2\pi)^4}
\frac{1}{(q^2-M_{D_i}^2+i\epsilon)((P-q)^2-M_{\phi_i}^2+i\epsilon)}\ ,\qquad
s\equiv P^2\ .
\eea
Note that $g(s)_i$ is counted as $\mathcal{O}(p)$ and its explicit expression is
\bea
g(s)_i\al=\al\frac{1}{16\pi^2}\bigg\{{a}(\mu)+\ln\frac{M_{D_i}^2}{\mu^2}+\frac{s-M_{D_i}^2+M_{\phi_i}^2}{2s}\ln\frac{M_{\phi_i}^2}{M_{D_i}^2}\nonumber\\
\al\al+\frac{\sigma_i}{2s}\big[\ln(s-M_{\phi_i}^2+M_{D_i}^2+\sigma_i)-\ln(-s+M_{\phi_i}^2-M_{D_i}^2+\sigma_i)\nonumber\\
\al\al+\ln(s+M_{\phi_i}^2-M_{D_i}^2+\sigma_i)-\ln(-s-M_{\phi_i}^2+M_{D_i}^2+\sigma_i)\big]\bigg\}\ ,
\label{eq:g}
\eea
with $\sigma_i=\{[s-(M_{\phi_i}+M_{D_i})^2][s-(M_{\phi_i}-M_{D_i})^2]\}^{1/2}$
and $\mu$ the renormalization scale. One can define a $\mu$-independent parameter
$\tilde{a}\equiv{a}(\mu)+\ln({M_{D_i}^2}/{\mu^2})$, since a change of $\mu$
in the logarithm can be compensated by $a(\mu)$. Notice that the parameter
$\tilde{a}$ in $g(s)$ of Eq.~\eqref{UniUChPT} cannot be absorbed by redefining
the LECs. It is introduced through the dispersion integral along the right-hand
cut, and is a free parameter in principle. The only constraint here is from
the requirement of a proper power counting: while all other terms in
Eq.~\eqref{eq:g} are of order $\order{p}$, $\tilde a$ should be much smaller
than 1 so that its presence will not cause a breaking of the power counting if we
expand the resummed amplitude to a certain order, i.e. $\tilde a = \order{p}$.
The kernel matrix
$\mathcal{T}(s)$ can be obtained perturbatively by matching to the ChPT
amplitudes order by order. Up to NNLO, it can be expressed as
\bea
\label{eq:Tuni}
\mathcal{T}(s)=\mathcal{A}^{(1)}(s)+\mathcal{A}^{(2)}(s)+\mathcal{A}^{(3)}(s)-
\mathcal{A}^{(1)}(s)\cdot g(s)\cdot \mathcal{A}^{(1)}(s),
\eea
where $\mathcal{A}^{(n)}(s)$ ($n=1,2,3$) stand for the partial wave amplitudes
from the perturbative calculation with the superscript $n$ denoting the chiral
dimension. Notice that the right hand cut from the NNLO amplitude is subtracted
in the last term in order to avoid  double counting in the unitarization. In the
function $g(s)$ in the above equation, the subtraction constant $\tilde a$ may
be removed as it can be absorbed into the redefinition of the LECs in
$\mathcal{A}^{(2)}(s)$.

The other approach is the so-called inverse amplitude method
(IAM)~\cite{Truong:1988zp,Dobado:1989qm,Oller:1998hw}. In our case, the IAM
unitized amplitudes has the matrix form
\bea
T(s)=\tilde{T}^{(1)}(s)\cdot\left[\tilde{T}^{(1)}(s)-
\tilde{T}^{(2)}(s)\right]^{-1}\cdot \tilde{T}^{(1)}(s)\ ,\label{UniIAM}
\eea
where
\bea
\tilde{T}^{(1)}(s)\equiv\mathcal{A}^{(1)}(s)\ ,\quad \tilde{T}^{(2)}(s)
\equiv\mathcal{A}^{(2)}(s)+\mathcal{A}^{(3)}(s)\ .
\eea
The above assignments guarantee that the unitarized amplitudes exactly obey
unitarity when the perturbatively unitary equations are employed, i.e.,
\bea
{\rm Im}\,\mathcal{A}^{(1)}(s)=0\ ,\quad {\rm Im}\,\mathcal{A}^{(2)}(s)=0\ ,
\quad {\rm Im}\,\mathcal{A}^{(3)}(s)=\mathcal{A}^{(1)}(s)\,\tilde{\rho}(s)\,
\mathcal{A}^{(1)}(s)^\dagger\ ,
\eea
with $\tilde{\rho}(s)={\rm diag}\{\tilde{\rho}(s)_i\}$,
$\tilde{\rho}(s)_i=-{q_i}/{(8\pi \sqrt{s})}$ and $q_i$ is the magnitude of the
center-of-mass (CM) three-momentum in the $i^{\rm th}$ channel.

\section{Calculation of the scattering lengths\label{SecSC}}

\subsection{Definition and pion mass dependence}

Given definite strangeness $S$ and isospin $I$, the $S$-wave scattering lengths
of the $i^\text{th}$ channel are related to the diagonal elements of the
$T$-matrix,~\footnote{We are using the sign convention such that the scattering
length for a repulsive interaction is negative.}
\bea
a^{(S,I)}_i=
-\frac{1}{8\pi(M_{D_i}+M_{\phi_i})}T^{(S,I)}_{\ell=0}(s_\text{th})_{ii}\ ,
\qquad s_\text{th}=(M_{D_i}+M_{\phi_i})^2 \ .
\eea
Here, $M_{D_i}$ and $M_{\phi_i}$ denote the masses of the charmed meson and
Goldstone boson $\phi$ in the channel $i$, respectively, and
$T^{(S,I)}_{\ell=0}(s_\text{th})$ stands for the $S$-wave unitarized amplitude
at threshold using either UChPT given by Eq.~(\ref{UniUChPT}) or IAM given by
Eq.~(\ref{UniIAM}).

Due to the short lifetime of the charmed meson, there are no experimental data
for $D$-$\phi$ scattering lengths. Nevertheless, lattice QCD calculations in
the last a few years provide very valuable information on the interaction
between the charmed mesons and light pseudoscalar
mesons~\cite{Liu:2008rza,Liu:2012zya,Mohler:2012na,Mohler:2013rwa}. Since the
lattice calculations were performed at several unphysical pion masses, in order
to describe these lattice data, one should know the pion mass dependence of the
scattering lengths. This is achieved by replacing all the quantities in the
expressions by the pion mass dependent ones. For the involved meson masses, we
have
\bea
M_K=\sqrt{\mathring{M}_K^2+M_\pi^2/2 }
\ ,\quad
M_D=\mathring{M}_D+(h_1+2h_0)\frac{M_\pi^2}{\mathring{M}_D}\ ,\quad
M_{D_s}=\mathring{M}_{D_s}+2h_0\frac{M_\pi^2}{\mathring{M}_{D_s}}\
.\label{massExtrapolation}
\eea
Note that all the formulae shown above are of NLO for the pion
mass dependence.~\footnote{
In Eq.~(\ref{massExtrapolation}),  although the formula we used for the kaon
mass is a LO expression in SU(3) ChPT, it contains two parts: the part
$\sim \mathring{M}_K^2$ proportional to $B_0 m_s$ remains in the SU(2) chiral limit
and is regarded as a LO contribution of the pion mass dependence, while the part
related to $B_0m_{u/d} \sim M_\pi^2/2$ vanishes in the SU(2) chiral limit and is thus
a NLO contribution. In this sense, we spelled out the pion mass dependence
for all of the masses and decay constants consistently up to the order
$M_\pi^2$.
}
Here the LEC $h_1$ can be fixed by the mass difference between $D$ and $D_s$.
Using these
two equations, one has
\cite{Guo:2009ct}
\bea
h_1=\frac{M_{D_s}^2-M_D^2}{4(M_K^2-M_\pi^2)}=0.4266\ ,
\eea
where the physical values for the meson masses are used, i.e.,
$M_\pi=138$~MeV, $M_K=496$~MeV, $M_D=1867$~MeV and
$M_{D_s}=1968$~MeV.\footnote{The mass of the $\eta$ is always expressed in
terms of $M_\pi$ and $M_K$ through the Gell-Mann--Okubo mass relation,
$3\,M_\eta^2=4M_K^2-M_\pi^2$. } The pion decay constant should also be
substituted by~\cite{Gasser:1984gg}
\be
F^{}_\pi=F^{}_0\left\{1- 2 \mu^{}_\pi - \mu^{}_K +
\frac{4M_\pi^2}{F^2_0}\left[L_4^r(\mu)+L_5^r(\mu)\right]+
\frac{8M_K^2}{F_0^2}L_4^r(\mu)\right\}\ ,
\ee
where $M_K$ is understood as the
one in Eq.~(\ref{massExtrapolation}), and $\mu^{}_\phi$ is a scale dependent
function for the Goldstone boson $\phi$
\be
\mu^{}_\phi = \frac{M_\phi^2}{32\pi^2 F_0^2} \ln \frac{M_\phi^2}{\mu^2}.
\ee
So far, except for $h_1$, the LECs $L_4^r$, $L_5^r$ and $h_0$ and the chiral
limit quantities $\mathring{M}_K$, $\mathring{M}_D$, $\mathring{M}_{D_s}$ and
$F_0$ are all unknown.
Since we will fit to the lattice results on the scattering lengths calculated in
Ref.~\cite{Liu:2012zya}, we choose to fix the above mentioned quantities from
fitting to the lattice data calculated using the same gauge configurations. In
addition, the kaon decay constant $F_K$ data are also included to have a bigger
data set for fixing $F_0$ and $L_{4,5}^r$. The pion mass dependence of $F_K$ is
given by~\cite{Gasser:1984gg}
\bea
F_K =
F_0\bigg\{1 - \frac34 \left( \mu_\pi + 2 \mu_K + \mu_\eta \right)
+ \frac{4M_\pi^2}{F_0^2}L_4^r(\mu) +
\frac{4M_{K}^2}{F_0^2}\left[2L_4^r(\mu)+L_5^r(\mu)\right]\bigg\},
\eea
with $M_\eta^2=(4M_K^2-M_\pi^2)/3$ and $M_K$ given by
Eq.~(\ref{massExtrapolation}).

\begin{figure}[t]
\begin{center}
\epsfig{file=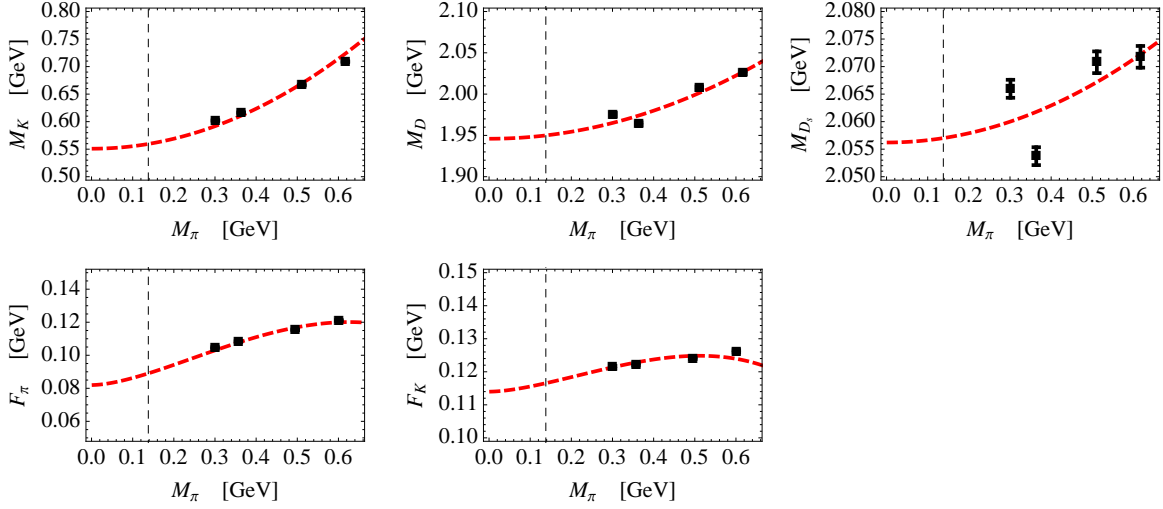,scale=0.7}
\end{center}
\caption{Chiral extrapolation of masses and decay constants. All the lattice data
are obtained from the same ensembles, namely M007-M030. Data for $M_K$, $M_D$
and $M_{D_s}$ is taken from Ref.~\cite{Liu:2012zya} and the one for $F_\pi$ and
$F_K$ from Ref.~\cite{WalkerLoud:2008bp}. Except for $M_{D_s}$, the data errors
are so tiny that we do not show them explicitly in the plots. The vertical
dashed line corresponds to the physical pion mass.}\label{ChiralExtra}
\end{figure}

The lattice data for $M_K$, $M_D$ and $M_{D_s}$ are taken from
Ref.~\cite{Liu:2012zya}. There are four data sets for each quantity,
corresponding to the four ensembles (labelled by M007, M010, M020 and M030) with
pion mass approximately 301.1~MeV, 363.8~MeV, 511.0~MeV and 617.0~MeV, in order.
Since the same ensembles are employed in
Ref.~\cite{WalkerLoud:2008bp}, we take the data for $f_\pi$
and $f_K$ from Ref.~\cite{WalkerLoud:2008bp}, where $f_\pi=\sqrt{2}F_\pi$ and
$f_K=\sqrt{2}F_K$.
Those lattice data are well described as shown in
Fig.~\ref{ChiralExtra}~\footnote{We have neglected the subtleties
due to the use of mixed action gauge configurations in the lattice calculations,
which in principle requires to use the partially quenched ChPT instead of the
standard one for the chiral extrapolation, and the effect of finite lattice
spacing, see Ref.~\cite{WalkerLoud:2008bp}.} when the parameters take the values
given in Table~\ref{TabChiralExtraLECs}. Our fitting values for $L_{4,5}^r$ are
consistent with the determinations given in
Refs.~\cite{Bazavov:2009fk,Dowdall:2013rya}. Therein, the values are obtained
at $\mu=M_\eta$, and the corresponding values transformed to $\mu=M_\rho$ can
be found in Ref.~\cite{Bijnens:2014lea}.

\begin{table}[bt]
\caption{Parameters for chiral extrapolation. $L_4^r$ and $L_5^r$ are obtained
at $\mu=M_\rho$ (= 775.5~MeV). The masses and decay constant in the chiral limit
are in units of MeV. $h_0$ and $h_1$ and dimensionless. The asterisk marks an
input value.}\label{TabChiralExtraLECs}
\vspace{-0.5cm}
\bea
\begin{array}{cccccccccc}
\hline\hline
 \mathring{M}_K &\mathring{M}_D    &\mathring{M}_{D_s}&h_{0}&h_1&F_0&10^{3}\cdot L_4^r& 10^{3}\cdot
 L_5^r\\
\hline
560.41&1940.4&2061.2&0.0172&0.4266^\ast&73.31&0.0095&1.3264\\
\hline\hline
\end{array}\nonumber
\eea
\end{table}

\subsection{Fits to lattice data on the scattering lengths}
\subsubsection{Introduction to the fitting procedure}

Since all the necessary preparations are completed, we proceed to the
description of the lattice QCD data of the $S$-wave scattering lengths. There
are two points to be discussed before carrying out the fits.

The first one is related to the lattice data. From Ref.~\cite{Liu:2012zya}, 20
data for 5 channels are available.  Amongst the five channels, the $D_s\pi$ with
$(S,I)=(1,1)$ can actually be coupled to the isovector $D K$ channel while the
other four are single channels. Although in Ref.~\cite{Liu:2012zya} only the
$D_s\pi$ interpolating operator  was constructed and used, the
propagation of all the quarks should know about the presence of the coupled $DK$
channel with $(S,I)=(1,1)$ because the channel-coupling in this case does not
require disconnected Wick contractions which were not included in
Ref.~\cite{Liu:2012zya}. Thus, we will describe the $D_s\pi$ data using a
coupled-channel unitarized amplitude.

In addition, lattice QCD results were published in the last two years for two
more channels: $D\pi$ with $(S,I)=(0,1/2)$~\cite{Mohler:2012na} and $DK$ with
$(S,I)=(1,0)$~\cite{Mohler:2013rwa}. These channels are more difficult since
both of them involve disconnected Wick contractions,~\footnote{It is shown in
Ref.~\cite{Guo:2013nja} that as long as the singly disconnected Wick
contractions contribute, which is the case for the isoscalar $DK$ channel, they
are of LO in both the $1/N_c$ and chiral expansion. Therefore, they cannot be
neglected.} but they are also more interesting as they can provide valuable
information for the lightest scalar charmed mesons in the corresponding
channels. The calculation for the $D\pi$ scattering was performed using $N_f=2$
gauge configurations, and the $DK$ calculation has results from both $N_f=2$ and
$N_f=2+1$ gauge configurations.
Because the amplitudes derived here are based on SU(3) ChPT, we will only
include in the fits the new result with $N_f=2+1$, i.e. $a_{DK\to
DK}^{(1,0)}=-1.33(20)$~fm obtained at $M_\pi=156$~MeV, and the unitarized
amplitude used in the fits is obtained including the $D_s\eta$ coupled channel.
Notice that these new lattice calculations use gauge configurations and actions
different from those in Ref.~\cite{Liu:2012zya}, the chiral-limit masses for the
kaon and charmed mesons should take different values from those given in
Table~\ref{TabChiralExtraLECs}. Because the physical masses of the involved
ground state mesons such as the kaon and charmed mesons were reproduced rather
well with the lattice setup  used in Ref.~\cite{Mohler:2013rwa} (for details, see
Ref.~\cite{Lang:2014yfa}), the chiral-limit values of the involved meson masses
and $F_0$ are determined by requiring them to coincide with the corresponding
phyiscal values at the physical pion mass, namely, $\mathring{M}_K=486.3$~MeV,
$\mathring{M}_D=1862.3$~MeV, $\mathring{M}_{D_s}=1967.7$~MeV and
$F_0=76.23$~MeV. The values for the LECs in the extrapolating expressions of
these quantities are the same as those listed in Table~\ref{TabChiralExtraLECs}.

The other point concerns the LECs to be determined. There are 7 unknown
LECs in total:  $h_2$, $h_3$, $h_4$, $h_5$, $g_1$, $g_2$ and $g_3$. As mentioned
in the previous work~\cite{Liu:2012zya}, $h_2$~($h_3$) and $h_4$~($h_5$) are
largely correlated. Therefore, redefinitions of the LECs are employed to reduce
these correlations, which are
\bea
h_{24}=h_2+h_4^\prime\ ,\quad h_{35}=h_3+2\,h_5^\prime\ ,\quad h_4^\prime=h_4\bar{M}_D^2\ ,\quad h_5^\prime=h_5\bar{M}_D^2\ .
\eea
The new parameters $h_{24}$, $h_{35}$, $h_4^\prime$ and $h_5^\prime$ will be
determined in our fits. The average of the physical masses of the charmed $D$
and $D_s$ mesons, $\bar{M}_D=(M_D^\text{phy}+M_{D_s}^\text{phy})/2$, is
introduced to make the four new parameters dimensionless. Similarly, for the
LECs from the NNLO contact terms, $g_2$ and $g_3$ are largely correlated
with each other, and it is better to redefine these LECs as
\bea
g_{23}=g_2^\prime-2g_3^\prime\ ,\quad g_1^\prime=g_1\bar{M}_D\ ,\quad
g_2^\prime=g_2\bar{M}_D\ ,\quad g_3^\prime=g_3\bar{M}_D^3\ .
\eea
The parameters $g_1^\prime$, $g_{23}$ and $g_3^\prime$ have a dimension of
inverse mass and will be fixed from fitting to the lattice data. One
can fix $g_1'$ and $g_{23}$ separately only when the coupled-channel unitarized
amplitudes are used, i.e. from fitting to the lattice results of the $D_s\pi$
and the isoscalar $DK$ scattering lengths.
The single-channel unitarized amplitudes is only sensitive to the combination
$g_{123}=g_{23}-g_1^\prime$, instead of $g_1'$ and $g_{23}$ separately, and
$g_3'$.

\subsubsection{Results}

We will try different fit procedures. In the fit $\text{UChPT-6(a)}$, all of the
20 data points for 5 channels, with pion masses from 301~MeV up to 617~MeV, in
Ref.~\cite{Liu:2012zya} as well as the $N_f=2+1$ datum for the isoscalar $DK$
channel, with an almost physical pion mass of 156~MeV, in Ref.~\cite{Mohler:2013rwa}
are taken into consideration. We notice that there are two possibilities for a
scattering length to be negative in our sign convention: a repulsive
interaction, and an attractive interaction with a bound state pole below the
threshold. In the $(S,I)=(1,0)$ channel, there is the well-known state
$D_{s0}^*(2317)$ below the $DK$ threshold which was not included as an explicit
degree of freedom in our theory. Because the number of data is small but the
number of parameters is large, a direct fit to these lattice data might result
in solutions which are not physically acceptable. For instance, within the range
of the parameters of a direct fit, the $(S,I)=(1,0)$ $DK$ channel could even be
repulsive which is reflected by the fact that the kernel of the unitarized amplitude
takes a positive value at the threshold.
Given that the LO interaction in the corresponding $DK$ channel is the most
attractive one among all the charmed meson--Goldstone boson scattering
processes, see Table~II in Ref.~\cite{Liu:2012zya} for instance, we regard such
a situation as unacceptable. Therefore, we put a constraint by hand requiring
that when all the particles take their physical values there is a bound state
pole in the $(S,I)=(1,0)$ channel at 2317~MeV. Following
Ref.~\cite{Liu:2012zya}, this is done by adjusting the subtraction constant
$\tilde a$ in the loop function $g(s)$ in the unitarized amplitude,
Eq.~\eqref{UniUChPT},  to produce the pole at the right position. The resulting
values of the LECs from the fit are shown in Table~\ref{TabLECs}.

\begin{table}[hpbt]
	\caption{Values of the LECs from the 6-channel fits using the method of
	UChPT. 	The $h_i$'s are dimensionless, and the  $g_1'$,
	$g_{23}$ and $g_3'$ are in GeV$^{-1}$.  }
	\label{TabLECs}
	\vspace{-0.5cm}
	\bea
	{
		\begin{array}{crrrr}
			\hline\hline
			&\text{UChPT-6(a)}&\text{UChPT-6(b)}&\text{UChPT-6($a^\prime$)}&\text{UChPT-6($b^\prime$)}\\
			&	\text{no prior}		 &	\text{no prior}			&\text{with prior}&\text{with
			prior}\\
			\hline
			h_{24}		& 0.79_{-0.09}^{+0.10}	& 0.76_{-0.09}^{+0.10}	& 0.83_{-0.10}^{+0.11}	& 0.80_{-0.10}^{+0.10}\\
			h_{35}		& 0.73_{-0.38}^{+0.50}	& 0.81_{-0.62}^{+0.95} 	& 0.43_{-0.23}^{+0.23}	& 0.40_{-0.29}^{+0.33}\\
			h_4^\prime 	& -1.49_{-0.57}^{+0.55}	& -1.56_{-0.65}^{+0.61}	& -1.33_{-0.60}^{+0.60}	& -1.72_{-0.63}^{+0.64}\\
			h_5^\prime 	& -11.47_{-2.79}^{+2.24}	&-15.38_{-7.20}^{+4.81}	& -4.25_{-0.66}^{+0.65} 	& -2.60_{-0.87}^{+0.84}\\
			g_{1}^\prime& -1.66_{-1.59}^{+0.31} 	& -2.44_{-0.64}^{+0.57}	& -1.10_{-0.23}^{+0.18}	& -1.90_{-0.35}^{+0.58}\\
			g_{23} 		& -1.24_{-1.51}^{+0.28} 	& -2.00_{-0.51}^{+0.52}  & -0.70_{-0.24}^{+0.19}  & -1.48_{-0.37}^{+0.61}\\
			g_3^\prime 	& 2.12_{-0.45}^{+0.55} 	& 2.85_{-0.96}^{+1.41}	& 0.98_{-0.14}^{+0.15}  	& 0.58_{-0.19}^{+0.20}\\
			\hline
			\chi^2/{\rm d.o.f.}	&\frac{31.52}{21-7}=2.25 &\frac{13.43}{16-7}=1.49&\frac{77.72-23.34}{21-7}=3.88	&\frac{51.71-16.60}{16-7}=3.90	\\
			\hline
			\hline
		\end{array}\nonumber
	}
	\eea
\end{table}

\begin{figure}[htbp]
\begin{center}
\includegraphics[width=\textwidth]{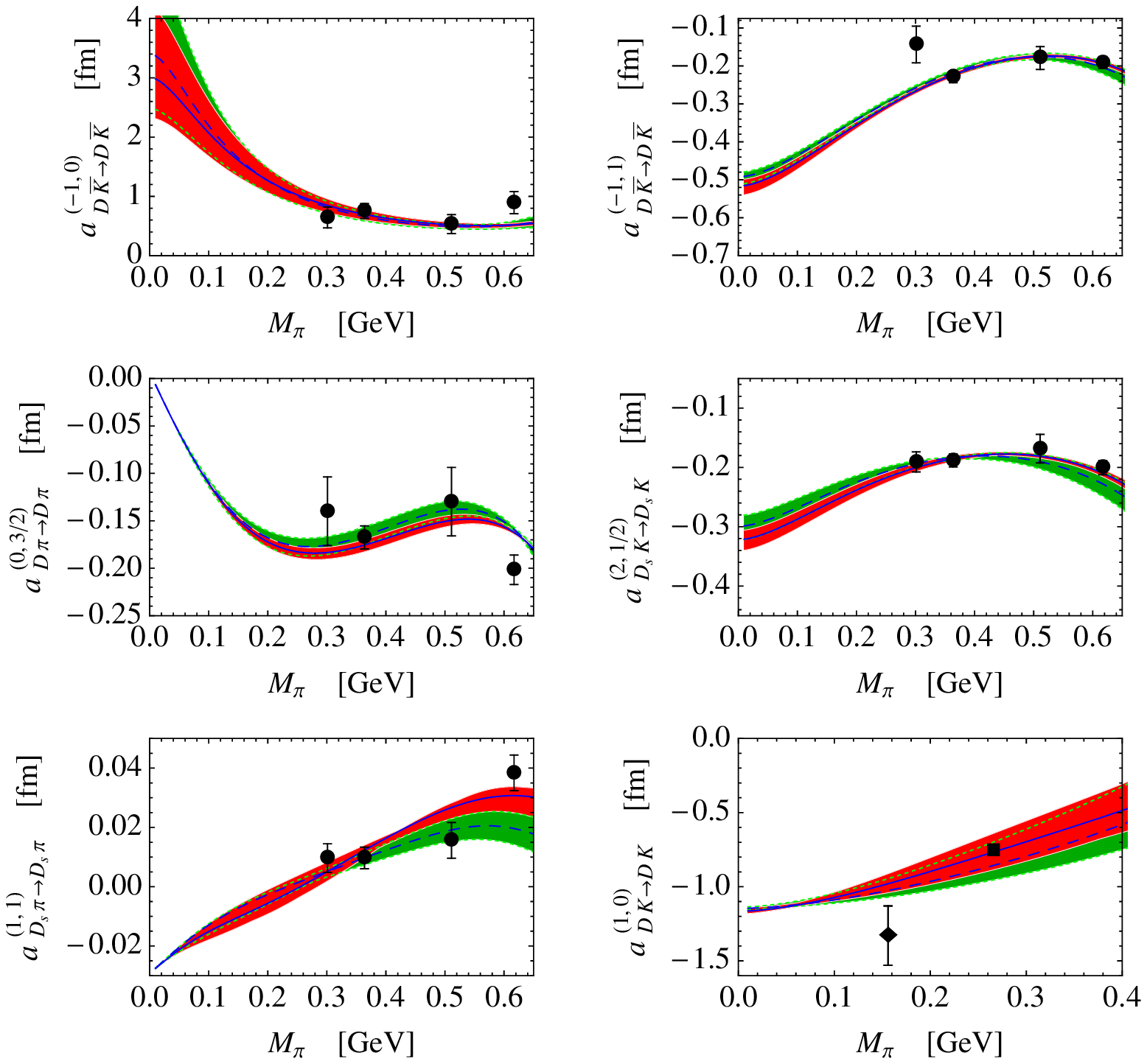}
\end{center}
\caption{Comparison of the results of the 6-channel fits (without a prior
$\chi^2$) to the lattice data of the scattering lengths.
U$\chi$PT-6(a): solid blue line with red band, U$\chi$PT-6(b): dashed blue line
with green band. The filled circles are lattice results in
Ref.~\cite{Liu:2012zya}, and the filled square (not included in the fits) and
diamond are taken from Ref.~\cite{Mohler:2013rwa}. }\label{figUChPT5}
\end{figure}

\begin{figure}[htbp]
	\begin{center}
		\includegraphics[width=\textwidth]{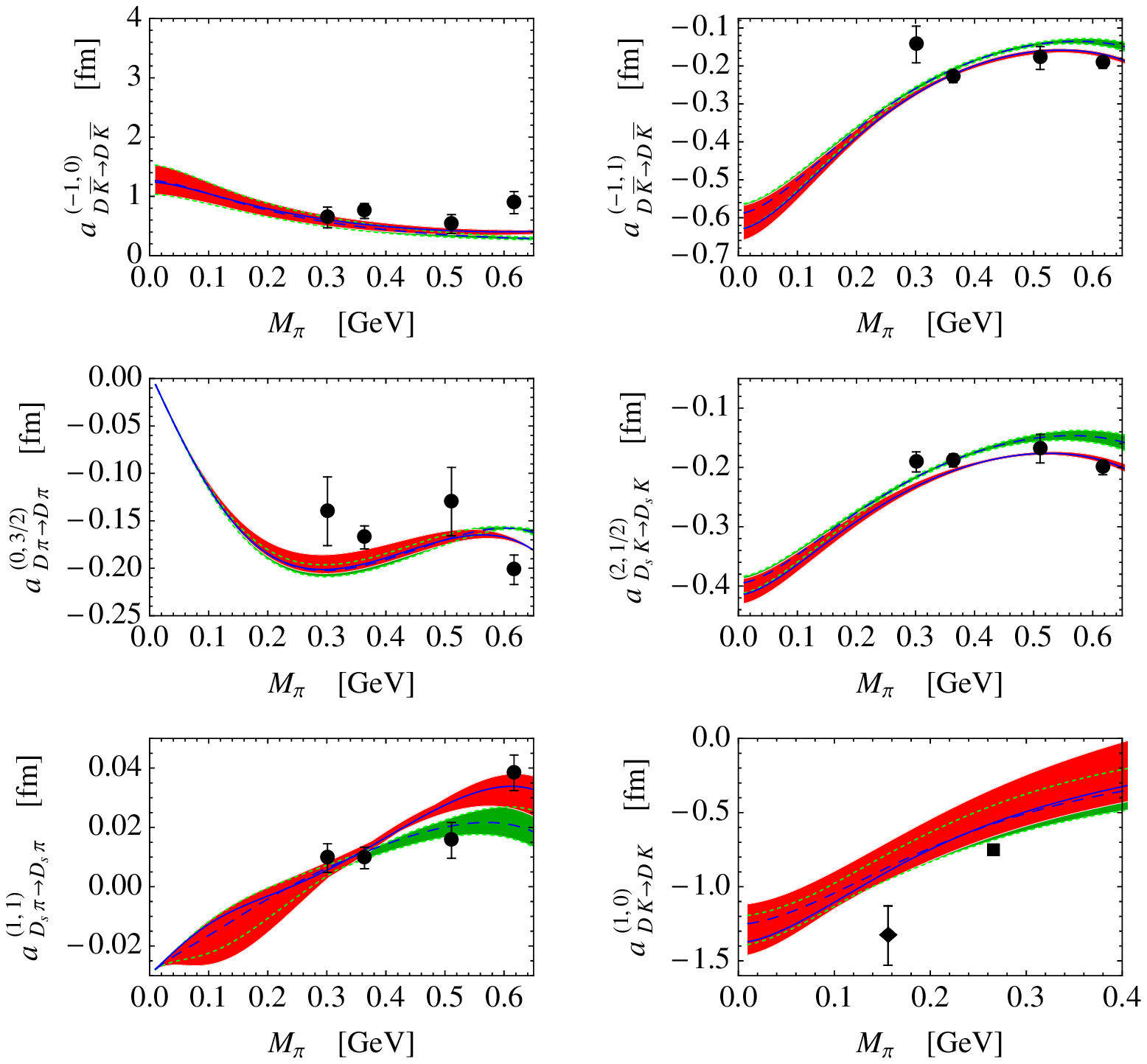}
	\end{center}
	\caption{Comparison of the results of the 6-channel fits (with a prior
	$\chi^2$) to the lattice data of the scattering lengths.
		U$\chi$PT-6(a): solid blue line with red band, U$\chi$PT-6(b): dashed blue line
		with green band. The filled circles are lattice results in
		Ref.~\cite{Liu:2012zya}, and the filled square (not included in the fits) and
		diamond are taken from Ref.~\cite{Mohler:2013rwa}. }\label{figUChPT6_prior}
\end{figure}

However, a pion mass larger than 600~MeV is definitely too large for
the chiral extrapolation using the standard ChPT. The unitarized approach
arguably has a larger convergence range than the standard ChPT. But the range is
not known a priori. Therefore, for the sake of comparison, we perform another
fit, denoted as UChPT-6(b), using the same method but excluding the lattice data
at $M_\pi=617$~MeV. The fit results are shown in the third column of
Table~\ref{TabLECs}. One can see that the values of all the LECs from these two
fits are similar, but those from UChPT-6(b) have larger uncertainties as a
result of being less constrained. The fit results from both fits are plotted in
Fig.~\ref{figUChPT5}. The bands represent the variation of the scattering
lengths with respect to the LECs within 1-$\sigma$ standard deviation. As we can
see, both fits describe the lattice data reasonably well with the exception that
the isoscalar $DK$ scattering length around $M_\pi=156$~MeV is too large in
comparison with the lattice result. However, both fits are consistent with
the $N_f=2$ lattice result for $DK$ at a pion mass around 266~MeV which was not
included in the fits.  We notice that the lattice ensemble for the
$M_\pi=156$~MeV datum has a rather small volume with $M_\pi L\approx 2.3$. It is
a bit too small for L\"uscher's finite volume formalism to be strictly
applicable, and thus this  datum might bear a large systematic uncertainty. The
isospin-3/2 $D\pi\to D\pi$ scattering length vanishes at the chiral limit as
required by chiral symmetry. Lattice discretization often breaks chiral
symmetry. However, due to the use of the domain-wall action for the valence
quarks in the lattice calculation of the pionic channels, the chiral behavior
is protected in our case. For related discussions in mixed-action ChPT, we
refer to Refs.~\cite{Bar:2003mh,Bar:2005tu,Tiburzi:2005is,Chen:2007ug}.

In both fits, the values of all the LECs except for $h_5'$ turn out to be of a
natural size.
However, the absolute value of the dimensionless LEC $h_5'$ is too large to be
natural.  This means that the absolute value of $h_5'$ is so
large that this single term would give a contribution larger than the LO
amplitude.
It would spoil the convergence, and thus the perturbative expansion, at
least for some quantities (although for some other quantities, due to
fine-tuned cancellation the sum of the NLO contribution could still be much
smaller than the LO one).  Therefore, we try to constrain all the LECs to
natural values following Ref.~\cite{Schindler:2008fh} which discusses the use of
the Bayesian method in effective field theories. Following
that paper, the so-called augmented chi-squared can be defined
by~\footnote{The method in Ref.~\cite{Schindler:2008fh} was only derived for the
case that the dependence on the parameters to be fitted is linear. Although our
case is non-linear and thus the augmented $\chi^2$ lacks a strict statistical
meaning, we still try this method as the $\chi^2$ defined in this way
comprises a "naturalness prior" so as to favor natural values for the LECs.
}
\bea
\chi_\text{aug}^2=\chi^2+\chi_\text{prior}^2\ ,
\label{eq:chisq}
\eea
where $\chi^2$ is the usual chi-squared used in the standard least
chi-squared fit and $\chi_\text{prior}^2$ is a prior chi-squared encoding the
naturalness requirement of the fit parameters. In our specific case, the
$\chi_\text{prior}^2$ is set to be the sum of squares of the fit LECs.
This means that we require the dimensionless LECs $h_i^{(')}$'s to be
$\mathcal{O}(1)$ and $g_i'$'s to be $\mathcal{O}(1~\text{GeV}^{-1})$. The
results by minimizing the augmented chi-squared are listed in the last two columns
in Table~\ref{TabLECs}, denoted as UChPT-6($a'$) and UChPT-6($b'$), where the
values for $\chi^2$ are given with $\chi^2_\text{prior}$ subtracted. One sees that
the value of $h_5'$ gets more natural at the price of a larger $\chi^2$.
A comparison of the scattering lengths with the lattice data in various channels
is given in Fig.~\ref{figUChPT6_prior}, and one can see that the lattice data in all six
channels can still be described reasonably well.

It turns out that in all of these fits $|h_5'|>|h_4'|$, which is consistent with
the $N_c$ counting $|h_4'|=\order{|h_5'|/N_c}$~\cite{Liu:2012zya}. The values of
the $h_i$'s are different from those obtained in Ref.~\cite{Liu:2012zya}. The
reason may be attributed to the use of the EOMS scheme in this work, and all of
$h_{2,3,4,5}$ absorb a power counting breaking contribution, see
Eq.~\eqref{EOMSren}.
For the case of the $D_s\pi$, the scattering length does not vanish at the limit
of a vanishing pion mass. This is due to the presence of the $DK$-loop in the
coupled-channel amplitude which has a nonvanishing contribution in the SU(2)
chiral limit. We have checked that the elastic contribution tends to zero as
$M_\pi$ approaches zero as required by  chiral symmetry.

\begin{table}[hbt]
	\caption{Values of the LECs from the 4-channel fits using both the methods
	of UChPT and IAM. 	The $h_i$'s are dimensionless, and the
	$g_{123}=g_{23}-g'$ and $g_3'$ are in GeV$^{-1}$. }
	\label{TabLECs4}
	\vspace{-0.5cm}
	\bea
	{
		\begin{array}{crr}
			\hline\hline
			&\text{UChPT-4}&\text{IAM-4}\\
			\hline
			h_{24}		& 0.50_{-0.10}^{+0.09}	& 0.53_{-0.07}^{+0.07}\\
			h_{35}		&-0.89_{-0.91}^{+0.93}	&-0.59_{-1.11}^{+1.04}\\
			h_4^\prime 	& 1.23_{-1.08}^{+1.03}	& 0.64_{-0.66}^{+0.66}\\
			h_5^\prime 	&-3.09_{-4.72}^{+4.69} 	&-6.08_{-5.99}^{+6.05}\\
			g_{123}& 0.18_{-0.18}^{+0.18} & 0.23_{-0.22}^{+0.21}\\
			g_3^\prime 	& 1.01_{-0.86}^{+0.87}  	& 1.42_{-1.10}^{+1.08}\\
			\hline
			\chi^2/{\rm d.o.f.}	&\frac{13.59}{16-6}=1.36	&\frac{13.97}{16-6}=1.40	\\
			\hline
			\hline
		\end{array}\nonumber
	}
	\eea
\end{table}

\begin{figure}[t]
\begin{center}
\includegraphics[width=\textwidth]{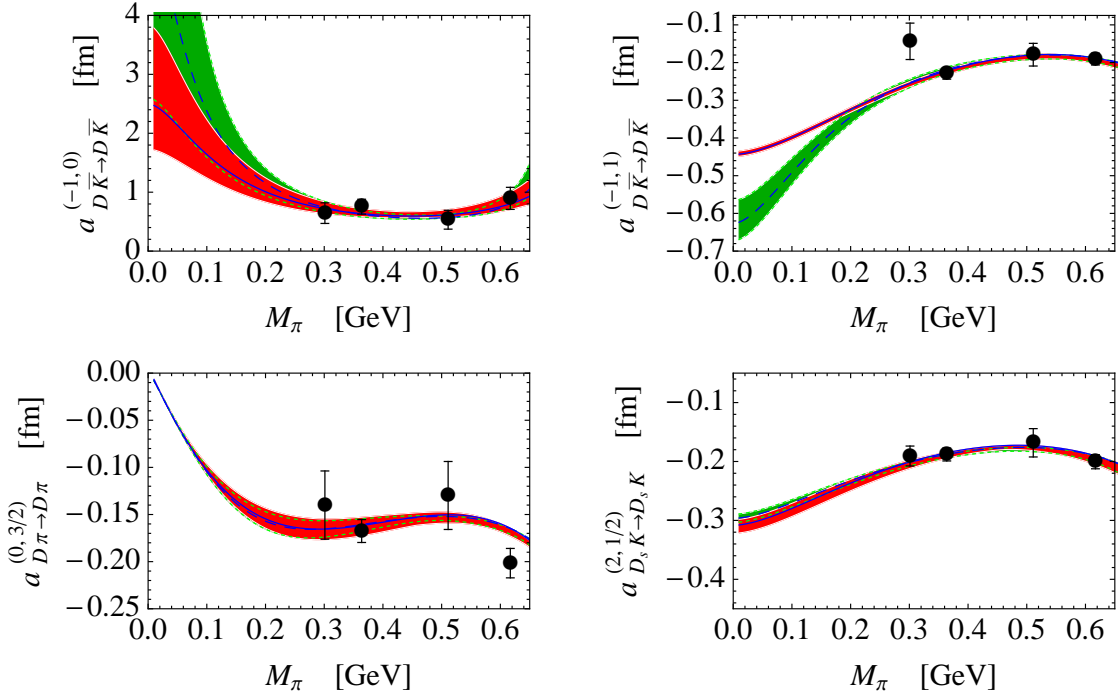}
\end{center}
\caption{
Comparison of the results of the 4-channel fits to the lattice data of
the scattering lengths.
U$\chi$PT-4: solid red line with blue band, IAM-4: dashed red line with green
band. The lattice data are taken from Ref.~\cite{Liu:2012zya}.}\label{figUChPT4}
\end{figure}

For comparison, we also perform fits  with just the four single-channel data,
i.e. the $D_s\pi$ and isoscalar $DK$ data are excluded. For this case, we use
two different unitarization methods: UChPT, to be denoted as UChPT-4, and IAM,
to be denoted as IAM-4. We did not use the IAM approach in the 6-channel fits
because this approach is not suitable to unitarize a perturbative amplitude with
a zero LO contribution. As can be seen from Eq.~\eqref{UniIAM}, if the LO
amplitude vanishes the unitarized one will vanish as well. This happens to the
case of the $D_s\pi$. The UChPT approach is free of this problem.
The results of these two fits are compiled in Table~\ref{TabLECs4}. Notice that
in this case $g_1'$ and $g_{23}$ cannot be determined separately, and the
effective combined parameter is $g_{123}=g_{23}-g_1'$. One sees that the values
of LECs from the fits using different unitarization methods are consistent with
each other,~\footnote{However, not all of the LECs in these different
unitarization methods ought to take the same values. One can see this by
expanding the IAM resummed amplitude up to $\order{p^3}$. Considering the single
channel case for simplicity, one has $T_\text{IAM}(s)=\mathcal{A}^{(1)}(s) +
\mathcal{A}^{(2)}(s) + \mathcal{A}^{(3)}(s) +
[\mathcal{A}^{(2)}(s)]^2/\mathcal{A}^{(1)}(s) + \order{p^4}$. It is different
from that of UChPT, $T_\text{UChPT}(s)=\mathcal{A}^{(1)}(s) +
\mathcal{A}^{(2)}(s) + \mathcal{A}^{(3)}(s) + \order{p^4}$. Thus, the LECs in
the $\order{p^3}$ Lagrangian could take different values. } but are only
marginally consistent with those in the 6-channel fits.
In addition, the uncertainties are quite large. More lattice simulations are
apparently necessary to pin down the LEC values. A comparison of the results of
the 4-channel fits to the lattice data in these channels are plotted in
Fig.~\ref{figUChPT4}.

\begin{table}[hpbt]
\centering
	\caption{Predictions of the scattering lengths at physical pion mass using
	the LECs determined in the 6-channel fits UChPT-6(b) and UChPT-6($b'$)
	in units of fm.
	}\label{TablePhy}
	\vspace{-0.5cm}
	\bea
	{
		\begin{array}{lcc}
			\hline\hline
				a^{(S,I)}_{ii}	&\text{UChPT-6(b)}&\text{UChPT-6($b'$)}\\
			\hline
			a^{(-1,0)}_{D\bar{K}\to D\bar{K}}			& 1.76_{-0.31}^{+0.39}		& 0.93_{-0.15}^{+0.15}\\
			a^{(-1,1)}_{D\bar{K}\to D\bar{K}}			&	-0.40_{-0.01}^{+0.01} 	&	-0.45_{-0.02}^{+0.01}\\
			a^{(0,\frac{1}{2})}_{D\pi\to D\pi} 	& 0.65_{-0.09}^{+0.11}	& 0.42_{-0.05}^{+0.04}		\\
			a^{(0,\frac{1}{2})}_{D\eta\to D\eta} 	&-0.18_{-0.04}^{+0.04}+ i\,0.00_{-0.00}^{+0.01}	&-0.21_{-0.04}^{+0.05}+ i\,0.01_{-0.01}^{+0.01}\\
			a^{(0,\frac{1}{2})}_{D_s\bar{K}\to D_s\bar{K}}&  -1.37_{-0.04}^{+0.21}+i\,0.61_{-0.02}^{+0.45} & -0.47_{-0.07}^{+0.06}+i\,0.50_{-0.16}^{+0.18} 		\\
			a^{(0,\frac{3}{2})}_{D\pi\to D\pi}		&  -0.14_{-0.01}^{+0.01}		& -0.15_{-0.01}^{+0.01}	\\
			a^{(1,0)}_{DK\to DK} 	& -1.04_{-0.03}^{+0.06}		& -1.50_{-0.26}^{+0.13}	\\
			a^{(1,0)}_{D_s\eta\to D_s\eta}	&-0.62_{-0.03}^{+0.02}+ i\,0.01_{-0.00}^{+0.01}		&-0.76_{-0.05}^{+0.05}+ i\,0.05_{-0.01}^{+0.00}	\\
			a^{(1,1)}_{D_s\pi\to D_s\pi} 	& -0.01_{-0.01}^{+0.01}			& -0.01_{-0.01}^{+0.01}	\\
			a^{(1,1)}_{DK\to DK}	&-1.11_{-0.09}^{+0.23}+i\,0.77_{-0.04}^{+0.27}	&-0.82_{-0.38}^{+0.59}+i\,1.64_{-0.11}^{+0.01}		\\
			a^{(2,\frac{1}{2})}_{D_sK\to D_sK}	&-0.25_{-0.02}^{+0.01}		&-0.32_{-0.01}^{+0.01}	\\
			\hline
			\hline
		\end{array}\nonumber
	}
	\eea
\end{table}

For reference, the values for the scattering lengths extrapolated to the
physical pion mass are presented in Table~\ref{TablePhy}.  The chiral limit
values in Table~\ref{TabChiralExtraLECs} are adopted for all the 16 channels
when performing the chiral extrapolation . Here we only show the results
using the 6-channel fits to the data with the pion mass up to 511~MeV, i.e.
UChPT-6(b) and UChPT-6($b'$). We notice that the numerical results of the scattering lengths extrapolated to the physical pion masses in some channels differ from those obtained in Ref.~\cite{Liu:2012zya}. This could indicate that the uncertainties are underestimated as the SU(3) formalism for UChPT was applied to pion masses higher than 500~MeV. We expect that the situation will improve when lattice results at lower pion masses are availble.

\subsubsection{Contribution of vector charmed mesons}

\begin{figure}[t]
\begin{center}
\epsfig{file=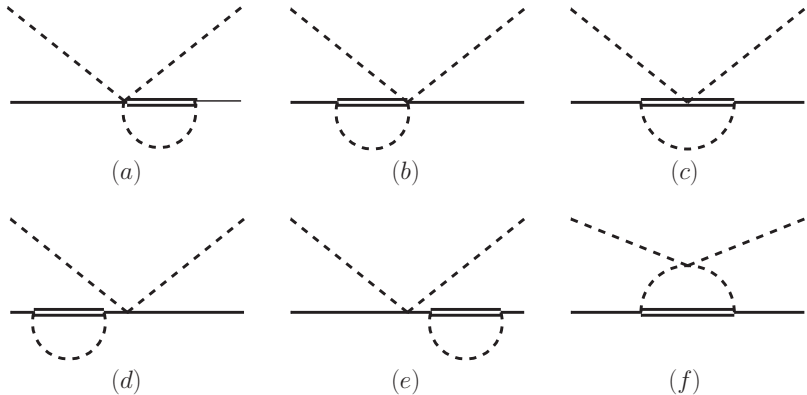,scale=0.55}
\caption{Feynman diagrams (including the vector charmed mesons) that survive in the heavy quark limit.}
\end{center}
\vspace{-5mm}
\label{FeynDiagsDstar}
\end{figure}

 In this section, contributions from vector charmed mesons will be
 included explicitly in order to quantify their influences on the $S$-wave
 scattering lengths. The diagrams that survive in the  heavy quark limit,  see
 also~Ref.~\cite{Geng:2010vw}, are taken into account and shown in
 Fig.~\ref{FeynDiagsDstar}.  Those diagrams vanishing in the heavy
 quark limit are suppressed by $1/m_c$
 and therefore are neglected.  We denote the vector charmed mesons by
 $D^\ast=(D^{\ast0},D^{\ast+},D_s^{\ast+})$, and the vertices
 involved in Fig.~\ref{FeynDiagsDstar} are described by  the following
 Lagrangian,
\bea
\mathcal{L}_{D^\ast D\Phi}=-\mathcal{D}_\mu D^{\ast\nu}\mathcal{D}^\mu D^{\ast\dagger}_{\nu}+M_0^\ast D^{\ast\nu} D^{\ast\dagger}_\nu+i\,\tilde{g}\left(D^\ast_{\mu}u^\mu D^\dagger-D\,u^\mu D^{\ast\dagger}_\mu\right)\ ,
\eea
where the covariant derivatives acting on $D^\ast$ are analogous to those defined in Eq.~(\ref{eq:cov}).
Further,  $M_0^\ast$ is the mass of  $D^\ast$ in the chiral limit.
The relation between the axial coupling constant $\tilde{g}$ defined here and
the coupling $g$ which is employed usually in the heavy meson
ChPT~\cite{Grinstein:1992qt,Wise:1992hn,Burdman:1992gh,Yan:1992gz} is
$\tilde{g}=\sqrt{M_DM_{D^\ast}}\,g$.  Following Ref.~\cite{Albaladejo:2015dsa},
we take $g=0.570\pm0.006$, determined by calculating the decay width of the
process $D^{\ast+}\to D^0\pi^+$, and then one gets $\tilde{g}\simeq
(1103.3\pm11.6)$~MeV. The calculations of the Feynman diagrams in
Fig.~\ref{FeynDiagsDstar} are straightforward but the analytical results are
too lengthy to be shown here. Similar to Eq.~(\ref{massExtrapolation}), the
pion-mass dependence of the $D^\ast$ and $D_s^\ast$ masses reads
\bea
M_{D^\ast}=\mathring{M}_{D^\ast}+(\tilde{h}_1+2\tilde{h}_0)\frac{M_\pi^2}{\mathring{M}_{D^\ast}}\ ,\quad
M_{{D_s}^\ast}=\mathring{M}_{{D_s}^\ast}+2\tilde{h}_0\frac{M_\pi^2}{\mathring{M}_{{D_s}^\ast}}\
,\label{massExtrapolationDstar}
\eea
where $\tilde h_0$ and $\tilde h_1$ are the analogues of $h_0$ and $h_1$,
respectively. In the heavy quark limit, one has $\tilde{h}_1={h}_1$
and $\tilde{h}_0={h}_0$. As discussed in Ref.~\cite{Altenbuchinger:2013vwa},
the breaking of heavy quark spin symmetry is only about $3\%$. Therefore, to a
good approximation, we impose these two heavy-quark limit relations.
The masses of the vector charmed mesons in the limit of $M_\pi\to 0$, i.e.
$\mathring{M}_{D^\ast}$ and $\mathring{M}_{{D_s}^\ast}$, are related to the
corresponding ones of the pseudoscalar charmed mesons via
\bea
{M_{D^\ast}}^\text{Phy.}-{M_D}^\text{Phy.}\simeq
\mathring{M}_{D^\ast}-\mathring{M}_D\ ,\qquad
{M_{{D_s}^\ast}}^\text{Phy.}-{M_{D_s}}^\text{Phy.}\simeq
\mathring{M}_{{D_s}^\ast}-\mathring{M}_{D_s}\ ,
\eea
with ${M_{D^\ast}}^\text{Phy.}=2008.6$~MeV and
${M_{D_s^\ast}}^\text{Phy.}=2112.3$~MeV, which are physical masses for $D^\ast$
and $D_s^\ast$, respectively. In parallel to the four kinds of 6-channel fits
in the previous section, we refit the $S$-wave scattering lengths and the
results are shown in Table~\ref{TabLECsDstar}. In each case, the LECs as well
as the chi-squared are almost same as before. This implies that the influence
of $D^\ast$ to the $S$-wave scattering lengths is marginal and it is a good
approximation to exclude them in the calculation.

\begin{table}[hpbt]
	\caption{Values of the LECs from the 6-channel fits (including explicit $D^\ast$) using the method of
	UChPT. 	The $h_i$'s are dimensionless, and the  $g_1'$,
	$g_{23}$ and $g_3'$ are in GeV$^{-1}$.  }
	\label{TabLECsDstar}
	\vspace{-0.5cm}
	\bea
	{
		\begin{array}{crrrr}
			\hline\hline
			&\text{UChPT-6(a)}&\text{UChPT-6(b)}&\text{UChPT-6($a^\prime$)}&\text{UChPT-6($b^\prime$)}\\
			&	\text{no prior}		 &	\text{no prior}			&\text{with prior}&\text{with
			prior}\\
			\hline
			h_{24}		& 0.80_{-0.08}^{+0.08}	& 0.80_{-0.09}^{+0.10}	& 0.85_{-0.10}^{+0.10}	& 0.85_{-0.10}^{+0.10}\\
			h_{35}		& 0.82_{-0.48}^{+0.60}	& 0.98_{-0.64}^{+0.97} 	& 0.50_{-0.23}^{+0.23}	& 0.59_{-0.29}^{+0.30}\\
			h_4^\prime 	& -1.27_{-0.51}^{+0.52}	& -1.40_{-0.62}^{+0.59}	& -1.22_{-0.57}^{+0.58}	& -1.59_{-0.61}^{+0.62}\\
			h_5^\prime 	& -11.61_{-3.07}^{+2.53}	&-15.06_{-7.31}^{+4.86}	& -3.87_{-0.69}^{+0.67} 	& -2.48_{-0.83}^{+0.84}\\
			g_{1}^\prime    & -2.94_{-0.36}^{+0.99} 	& -2.69_{-0.61}^{+0.51}	& -1.45_{-0.30}^{+0.20}	& -1.90_{-0.43}^{+0.39}\\
			g_{23} 		& -2.56_{-0.31}^{+0.99} 	& -2.28_{-0.48}^{+0.46}  & -1.10_{-0.31}^{+0.21}  & -1.51_{-0.45}^{+0.40}\\
			g_3^\prime 	& 2.15_{-0.49}^{+0.60} 	& 2.80_{-0.96}^{+1.42}	& 0.91_{-0.15}^{+0.15}  	& 0.56_{-0.19}^{+0.19}\\
			\hline
			\chi^2/{\rm d.o.f.}	&\frac{29.36}{21-7}=2.10 &\frac{13.75}{16-7}=1.53&\frac{74.22-21.56}{21-7}=3.76	&\frac{50.06-15.96}{16-7}=3.79	\\
			\hline
			\hline
		\end{array}\nonumber
	}
	\eea
\end{table}

\section{Summary and outlook \label{SecCon}}

We have computed the $D$-$\phi$ scattering amplitude that is valid up to the
NNLO in the chiral expansion within the framework of ChPT. The complete
analytical expressions for the amplitudes are given using a renormalization
procedure with the EOMS subtraction scheme. We show explicitly that
the UV divergences and the PCB terms, both of which stem from the loops, can be
absorbed into the LECs. We then obtained the EOMS-renormalized $D$-$\phi$
scattering amplitudes which are independent of the renormalization scale and
possess good properties such as correct power counting and proper analyticity.

In order to describe the lattice data on the $S$-wave scattering lengths at
relatively high pion masses and to account for the nonperturbative nature in the
channels like the $(S,I)=(1,0)$ $DK$, the aforementioned perturbative amplitudes
are inserted into a unitarization procedure to perform the chiral extrapolation
from large unphysical light quark masses down to the SU(2) chiral limit.
We tried different fitting procedures with and without a naturalness constraint.
It turns out that the absolute value of $h_5'$ could be quite large if the
naturalness constraint is not put by hand.
We want to stress that more lattice simulations in different channels
are necessary for a better determination of the involved LECs and a better
understanding of the scalar and axial-vector charmed mesons. When the
LECs are well constrained, we can make reliable predictions in the channels
which have not been calculated on the lattice and in the bottom sector
utilizing heavy quark spin symmetry.

\section*{Acknowledgements}

We would like to thank C. Hanhart and Tom Luu for useful discussions and
comments.
This work is supported in part by DFG and NSFC through funds provided to the Sino-German CRC 110 ``Symmetries and the Emergence of Structure in QCD'' (NSFC
Grant No. 11261130311),  and by NSFC (Grant No. 11165005). The work of UGM was supported in part by The Chinese Academy of Sciences (CAS)
President's International Fellowship Initiative (PIFI) grant no. 2015VMA076.

\bigskip

\appendix

\renewcommand{\theequation}{\thesection.\arabic{equation}}

\section{Definition of one-loop integrals\label{secloopint}}

\setcounter{equation}{0}

In this appendix, all the relevant one-loop integrals are defined. For the
current case, only one- and two-point loop functions are involved. As is
well known, each tensor one-loop integral can be expressed as a linear sum of
scalar one-loop integrals by using the method of Passarino--Veltmann (PV)
decomposition~\cite{Passarino:1978jh}. Hence, if the explicit expressions of the
scalar one-loop integrals are known, the loop amplitudes can be obtained
analytically.

Throughout this work, the ultraviolet divergence is contained in the quantity
$R$ which is defined by
\bea
R=\frac{2}{d-4}+\gamma_E-1-\ln(4\pi)\ ,
\eea
with $\gamma_E$
 the Euler constant and $d$ the space-time dimension. In addition,
we will denote the renormalization scale by $\mu$. In terms of these
notations, various loop integrals involved in the calculations are given as
follows:
\begin{itemize}
\item{One-point loop function:}
\bea
\tadpole_a=\frac{\mu^{4-d}}{i}\int\frac{\textrm{d}^dk}{(2\pi)^d}
\frac{1}{k^2-M_a^2+i0^+}=-\frac{M_a^2}{16\pi^2}
\left(R+\ln\frac{M_a^2}{\mu^2}\right) .
\eea
\item{Two-point loop function for unequal masses: ($M_a>M_b$)}
\bea
\al\al \{\mathcal{H}_{ab}(p^2),p^\mu \mathcal{H}_{ab}^{1}(p^2), g^{\mu\nu}
\mathcal{H}_{ab}^{00}(p^2)+p^\mu p^\nu \mathcal{H}_{ab}^{11}(p^2)\}
\nonumber\\
\al=\al
\frac{\mu^{4-d}}{i}\int\frac{\textrm{d}^dk}{(2\pi)^d} \frac{\{1,k^{\mu},k^{\mu}
k^{\nu}\}}{\left(k^2-M_a^2+i0^+\right)\left[(k+p)^2-M_b^2+i0^+\right]}\ ,
\eea
where the PV coefficients are given by
\bea
\mathcal{H}_{ab}^{1}(p^2)\al=\al
\frac{1}{2p^2}\left[\tadpole_a-\tadpole_b-(p^2+\Delta_{ab})\mathcal{H}_{ab}(p^2)
\right]\ ,\nonumber\\
\mathcal{H}_{ab}^{00}(p^2)\al=\al
\frac{1}{12p^2}\left\{(p^2+\Delta_{ab})\tadpole_a+(p^2-\Delta_{ab})\tadpole_b
+ \left[4p^2\,M_a^2-(p^2+\Delta_{ab})^2\right]\mathcal{H}_{ab}(p^2)\right\}
\nonumber\\
\al\al -\frac{1}{16\pi^2}\frac{1}{18}(p^2-3\Sigma_{ab})\ ,\nonumber\\
\mathcal{H}_{ab}^{11}(p^2)\al=\al
\frac{1}{3p^4}\left\{-(p^2+\Delta_{ab})\tadpole_a+
(2p^2+\Delta_{ab})\tadpole_b- \left[p^2\,
M_a^2-(p^2+\Delta_{ab})^{2}\right]\mathcal{H}_{ab}(p^2)
\right\}\nonumber\\
\al\al +\frac{1}{16\pi^2}\frac{1}{18p^2}(p^2-3\Sigma_{ab})\ ,\nonumber
\eea
 where we have defined $\Delta_{ab}
\equiv M_a^2-M_b^2$ and $\Sigma_{ab} \equiv M_a^2+M_b^2$.
The scalar two-point one-loop function $\mathcal{H}_{ab}(p^2)$ has the following
analytical form,
\bea
\mathcal{H}_{ab}(p^2) \al=\al
\frac{1}{16\pi^2}\left[ -R+1-\ln\frac{M_b^2}{\mu^2}+
\frac{\Delta_{ab}+p^2}{2\,p^2}\ln\frac{M_b^2}{M_a^2} \right. \nonumber\\
\al\al + \left.
\frac{p^2-(M_a-M_b)^2}{p^2}\rho_{ab}(p^2)
\ln\frac{\rho_{ab}(p^2)-1}{\rho_{ab}(p^2)+1}\right]\ ,
\eea
with
\bea
\rho_{ab}(p^2)\equiv\sqrt{\frac{p^2-(M_a+M_b)^2}{p^2-(M_a-M_b)^2}}.
\eea
To get the imaginary part above the threshold properly, one should take the
branch cut for the logarithm  along the negative real axis.
\item{Two-point loop function for equal masses:}
\bea
\al\al\{\mathcal{J}_{a}(p^2),p^\mu\mathcal{J}_{a}^{1}(p^2),g^{\mu\nu}
\mathcal{J}_{a}^{00}(p^2)+p^\mu p^\nu\mathcal{J}_{a}^{11}(p^2)\}\nonumber\\
\al=\al\frac{\mu^{4-d}}{i}\int\frac{\textrm{d}^dk}{(2\pi)^d}
\frac{\{1,k^{\mu},k^{\mu}
k^{\nu}\}}{\left(k^2-M_a^2+i0^+\right)\left[(k-p)^2-M_a^2+i0^+\right]}\
,\nonumber
\eea
where the Passarino--Veltmann coefficients are given by
\bea
\mathcal{J}_{a}^{1}(p^2)\al=\al -\frac{1}{2}\mathcal{J}_{a}(p^2)\
,\nonumber\\
\mathcal{J}_{a}^{00}(p^2)\al=\al
\frac{1}{12}(4M_a^2-p^2)\mathcal{J}_a(p^2)+\frac{1}{6}\tadpole_a+
\frac{1}{16\pi^2}\frac{1}{18}(6M_a^2-p^2)\ ,\nonumber\\
\mathcal{J}_{a}^{11}(p^2)\al=\al
\frac{1}{3p^2}\left[(p^2-M_a^2)\mathcal{J}_a(p^2)+\tadpole_a\right] +
\frac{1}{16\pi^2}\frac{1}{18p^2}(p^2-6M_a^2)\ .\nonumber
\eea
In this case, the scalar two-point one-loop function $\mathcal{J}_{a}(p^2)$ has
a much simpler analytical form,
\be
\mathcal{J}_{a}(p^2)=\frac{1}{16\pi^2}\left[
-R+1-\ln\frac{M_a^2}{\mu^2}+
\sigma_{a}(p^2)\ln\frac{\sigma_{a}(p^2)-1}{\sigma_{a}(p^2)+1}\right] ,\quad
\sigma_a(p^2)\equiv\sqrt{1-\frac{4M_a^2}{p^2}}.
\ee

\end{itemize}

\section{Loop amplitudes without explicit charmed vector mensons\label{secloopampl}}
\setcounter{equation}{0}

In order to express the loop amplitude in a short form, the following
abbreviation is adopted,
\bea
\mathcal{F}_{ab}^{(cd)}(s,t)\al=\al \left[3(s-M_a^2)+(s-M_c^2)\right]\tadpole_d
-\left(s-\Sigma_{bc}\right)^2\mathcal{H}_{cd}(s)+2 \left(t - 2 M_b^2\right)
\mathcal{H}^{00}_{cd}( s)\nonumber\\
\al\al + 2 \left(s - \Delta_{ab}\right) \left(s - \Sigma_{bc} \right)
\mathcal{H}^1_{cd}(s)-\left(s - \Delta_{ab}\right)^2 \mathcal{H}_{cd}^{11}( s)\ .
\label{eq:F}
\eea
We first list the loop amplitudes concerning the elastic scattering processes.
\begin{itemize}
\item{$D^0K^-\to D^0K^-$}
\bea
\mathcal{A}^{\rm loop}_{D^0K^-\to
D^0K^-}(s,t)\al=\al\frac{1}{16F^4}\bigg\{\mathcal{F}_{DK}^{(DK)}(s,t)+
2\mathcal{F}_{DK}^{(DK)}(u,t)+ \frac{3}{2}\mathcal{F}_{DK}^{(D_s\eta)}(u,t)+
\frac{1}{2}\mathcal{F}_{DK}^{(D_s\pi)}(u,t)\nonumber\\
\al\al +(s-u)\left(\tadpole_\eta+2\tadpole_K+\tadpole_\pi\right)-
4(s-u)\left[\mathcal{J}_\pi^{00}(t)+2\mathcal{J}_K^{00}(t)\right]\bigg\}.
\nonumber\\
\eea
\item{$D^+K^+\to D^+K^+$}
\be
\mathcal{A}^{\rm loop}_{D^+K^+\to D^+K^+}(s,t)=
\frac{1}{16F^4}\left\{\mathcal{F}_{DK}^{(D_s\pi)}(s,t)+
\mathcal{F}_{DK}^{(DK)}(u,t)-4(s-u)\left[\mathcal{J}_\pi^{00}(t)-
\mathcal{J}_K^{00}(t)\right]\right\}.
\ee
\item{$D^+\pi^+\to D^+\pi^+$}
\bea
\mathcal{A}^{\rm loop}_{D^+\pi^+\to D^+\pi^+}(s,t)\al=\al
\frac{1}{16F^4}\bigg\{\mathcal{F}_{D\pi}^{(D\pi)}(s,t)+
3\mathcal{F}_{D\pi}^{(D\pi)}(u,t)+
\mathcal{F}_{D\pi}^{(D_sK)}(u,t)\nonumber\\
\al\al+\frac{4}{3}(s-u)\left(2\tadpole_\pi+\tadpole_K\right)-
4(s-u)\left[2\mathcal{J}_\pi^{00}(t)+\mathcal{J}_K^{00}(t)\right]\bigg\}.
\eea
\item{$D^+\eta\to D^+\eta$}
\bea
\mathcal{A}^{\rm loop}_{D^+\eta\to D^+\eta}(s,t)=
\frac{1}{16F^4}\left[\frac{3}{2}\mathcal{F}_{D\eta}^{(D_sK)}(s,t)+
\frac{3}{2}\mathcal{F}_{D\eta}^{(D_sK)}(u,t)\right].
\eea
\item{$D_s^+K^+\to D_s^+K^+$}
\bea
\mathcal{A}^{\rm loop}_{D_s^+K^+\to D_s^+K^+}(s,t) \al=\al
\frac{1}{16F^4}\bigg[\mathcal{F}_{D_sK}^{(D_sK)}(s,t)+
\mathcal{F}_{D_sK}^{(D_sK)}(u,t)+\frac{3}{2}\mathcal{F}_{D_sK}^{(D\eta)}(u,t)+
\frac{3}{2}\mathcal{F}_{D_sK}^{(D\pi)}(u,t)\nonumber\\
\al\al
+(s-u)\left(\tadpole_\eta+2\tadpole_K+\tadpole_\pi\right)-
12(s-u)\mathcal{J}_K^{00}(t)\bigg].
\eea
\item{$D_s^+\eta\to D_s^+\eta$}
\bea
\mathcal{A}^{\rm loop}_{D_s^+\eta\to D_s^+\eta}(s,t)=
\frac{1}{16F^4}\left[3\,\mathcal{F}_{D_s\eta}^{(DK)}(s,t)
+3\,\mathcal{F}_{D_s\eta}^{(DK)}(u,t)\right].
\eea
\item{$D_s^+\pi^0\to D_s^+\pi^0$}
\bea
\mathcal{A}^{\rm loop}_{D_s^+\pi^0\to D_s^+\pi^0}(s,t) =
\frac{1}{16F^4}\left[\mathcal{F}_{D_s\pi}^{(DK)}(s,t)+
\mathcal{F}_{D_s\pi}^{(DK)}(u,t)\right].
\eea
\end{itemize}

As for the inelastic processes, the amplitudes become a little more complicated.
To reduce them, we further need
\bea
\mathcal{G}_{ab,cd}^{(ef)}(s,t)
= \frac{1}{2}\Delta_{bd}^2\mathcal{H}_{ef}(s)+
\frac{1}{2}\left(\Delta_{ac} - \Delta_{bd}\right)^2\mathcal{H}_{ef}^{11}(s)
- \Delta_{bd} \left(\Delta_{ac} - \Delta_{bd}\right)\mathcal{H}^1_{ef}(s)\
.\eea
In the above equation, the letters $a$ and $b$ ($c$ and $d$) label the incoming
(outgoing) particles, while $e$ and $f$ mark the particles in the loop. This
convention also holds for the abbreviations
$^1\mathcal{K}_{ab,cd}^{(ef)}(s,t)$ and $^2\mathcal{K}_{ab,cd}^{(ef)}(s,t)$,
whose explicit expressions are given by
\bea
^1\mathcal{K}_{ab,cd}^{(ef)}(s,t)\al=\al
\Delta_{ac}\left\{\frac{1}{2}\tadpole_f +
\frac{1}{2}t\left[\mathcal{H}_{ef}(t)+\mathcal{H}^{1}_{ef}(t)\right] -
\mathcal{H}^{00}_{ef}(t) - t\mathcal{H}^{11}_{ef}(t) \right\} ,\nonumber\\
^2\mathcal{K}_{ab,cd}^{(ef)}(s,t) \al=\al
-3(s-u)\mathcal{H}^{00}_{ef}(t)
-\Delta_{ac}\bigg[ \frac{1}{6}(6\Sigma_{de}+ 6\Delta_{df} - 13\Delta_{bd})
\mathcal{H}^1_{ef}(t)\nonumber\\
\al\al +
\frac{1}{6}(3\Sigma_{de} + 3\Delta_{df} - 2\Delta_{bd})
\mathcal{H}_{ef}(t) -3\Delta_{bd}\mathcal{H}_{ef}^{11}(t)\bigg].
\eea
In combination with the abovementioned notations, the inelastic one-loop
scattering amplitudes are given as follows:
\begin{itemize}
\item{$D^0\eta\to D^0\pi^0$}
\be
\mathcal{A}^{\rm loop}_{D^0\eta\to D^0\pi^0}(s,t)=
\frac{\sqrt{3}}{64F^4}\left[\mathcal{F}_{D\eta}^{(D_sK)}(s,t)+
\mathcal{F}_{D\pi}^{(D_sK)}(s,t)+
2\,\mathcal{G}_{D\eta,D\pi}^{(D_sK)}(s,t) + (s\leftrightarrow u)\right].
\ee
\item{$D_s^+K^-\to D^0\pi^0$}
\bea
\mathcal{A}^{\rm loop}_{D_s^+K^-\to D^0\pi^0}(s,t)\al=\al
\frac{\sqrt{2}}{16F^4}\bigg\{\frac{1}{2}\left[\mathcal{F}_{D_sK}^{(D\pi)}(s,t)+
\mathcal{F}_{D\pi}^{(D\pi)}(s,t)+
2\,\mathcal{G}_{D_sK,D\pi}^{(D\pi)}(s,t)\right]\nonumber\\
\al\al
+\frac{1}{4}\left[\mathcal{F}_{D_sK}^{(D_sK)}(s,t)+
\mathcal{F}_{D\pi}^{(D_sK)}(s,t)+
2\,\mathcal{G}_{D_sK,D\pi}^{(D_sK)}(s,t)\right] \nonumber\\
\al\al  -
\frac{s-u}{12}\left(3\tadpole_\eta+{11}\tadpole_\pi+{10}\tadpole_K\right)
+\frac{\Delta_{D_s D}}{24}
\left(3\tadpole_\eta-{5}\tadpole_\pi+{2}\tadpole_K\right)\nonumber\\
\al\al -\left(^1\mathcal{K}_{D_sK,D\pi}^{(\eta K)}(s,t)+
\,^2\mathcal{K}_{D_sK,D\pi}^{(\eta K)}(s,t)\right) \nonumber\\
\al\al -
\left(\frac{5}{3}\,^1\mathcal{K}_{D_sK,D\pi}^{(K\pi)}(s,t)+
\,^2\mathcal{K}_{D_sK,D\pi}^{(K\pi)}(s,t)\right) \bigg\}\ .
\eea
\item{$D_s^+K^-\to D^0\eta$}
\bea
\mathcal{A}^{\rm loop}_{D_s^+K^-\to D^0\eta}(s,t)\al=\al
\frac{\sqrt{6}}{16F^4}\bigg\{\frac{1}{4}\left(\mathcal{F}_{D_sK}^{(D_sK)}(s,t)+
\mathcal{F}_{D\eta}^{(D_sK)}(s,t)+
2\,\mathcal{G}_{D_sK,D\eta}^{(D_sK)}(s,t)\right) \nonumber\\
\al\al-\frac{1}{2}\left(\mathcal{F}_{D_sK}^{(DK)}(u,t)+
\mathcal{F}_{D\eta}^{(DK)}(u,t)+\mathcal{G}_{D_sK,D\eta}^{(DK)}(u,t)+
\mathcal{G}_{D_s\eta,DK}^{(DK)}(u,t)\right)\nonumber\\
\al\al+\left(^1\mathcal{K}_{D_sK,D\eta}^{(K\pi)}(s,t)-
\,^2\mathcal{K}_{D_sK,D\eta}^{(K\pi)}(s,t)\right) \nonumber\\
\al\al - \left(^1\mathcal{K}_{D_sK,D\eta}^{(\eta K)}(s,t)+
\,^2\mathcal{K}_{D_sK,D\eta}^{(\eta K)}(s,t)\right)\nonumber\\
\al\al +\frac{\Delta_{D_s D}}{6} (5M_\eta^2+8M_K^2-M_\pi^2)
\left(2\mathcal{H}_{K\pi}^1(t)+\mathcal{H}_{K\pi}(t)\right)\nonumber\\
\al\al - \frac{\Delta_{D_s D}}{3} (M_\eta^2-4M_K^2+M_\pi^2)
\left(2\mathcal{H}_{\eta K}^1(t)+\mathcal{H}_{\eta K}(t)\right)\nonumber\\
\al\al + \frac{\Delta_{D_s D}}{8}
\left(\tadpole_\eta+\tadpole_\pi-2\tadpole_K\right)-
\frac{s-u}{4}\left(\tadpole_\eta+\tadpole_\pi+6\tadpole_K\right)\bigg\}\ .
\eea
\end{itemize}

\section{Infrared regular parts of the loop integrals\label{secRegular}}
\setcounter{equation}{0}

The following expressions for the infrared regular parts are taken from
Ref.~\cite{Chen:2012nx} with the nucleon mass (pion) mass) replaced by the $D$ meson
(Goldstone boson) mass:
\begin{itemize}
\item{one-point: $a\in\{D,\,D_s\}$ }
\bea
\tadpole^{\rm reg.}_a=
     -\frac{M_a^2}{16\pi^2}\ln\frac{M_a^2}{\mu^2}\ .
\eea
\item{two-point: $a\in\{D,\,D_s\}$ and $b\in\{\pi,\,K,\,\eta\}$}

\bea
\mathcal{H}_{ab}^{\rm reg.}(s)\al=\al\frac{1}{16\pi^2}
\left(1-\log\frac{M_a^2}{\mu^2}\right)-
\frac{s-M_a^2}{2M_a^2}\frac{1}{16\pi^2}\left(1-\log\frac{M_a^2}{\mu^2}\right)
\nonumber\\
\al\al +
\frac{1}{32\pi^2}\left[\frac{M_b^2}{M_a^2}\left(3+\log\frac{M_a^2}{\mu^2}
\right)- \left(\frac{s-M_a^2}{M_a^2}\right)^2\log\frac{M_a^2}{\mu^2}\right]
+\mathcal{O}(p^3)\ .
\eea
\end{itemize}
The power counting breaking term of $\mathcal{F}_{ab}^{(cd)}(s,t)$ is of
$\mathcal{O}(p^2)$ and its explicit form reads
\bea
\mathcal{F}_{ab}^{(cd)}(s,t)^{\rm PCB}\al=\al\frac{1}{16\pi^2}
\left\{2(s-M_a^2)(s-M_c^2)\left[\frac{1}{2}\log\frac{M_c^2}{\mu^2}-1\right]-(s-M_a^2)^2\left[\frac{8}{9}-\frac{1}{3}\log\frac{M_c^2}{\mu^2}\right]\right.\nonumber\\
\al\al\left.-(s-M_c)^2\left[1-\log\frac{M_c^2}{\mu^2}\right]+2(t-2M_b^2)M_c^2\left[\frac{1}{9}-\frac{1}{6}\log\frac{M_c^2}{\mu^2}\right]\right\}\ .
\label{PCBmathF}
\eea
Since the difference between
$M_a^2$ and $M_c^2$ is at least $\mathcal{O}(p^2)$, the above expression can
be reduced to a simpler form
\bea
\mathcal{F}_{ab}^{(cd)}(s,t)^{\rm PCB}\al=\al\frac{1}{144\pi^2}
\left\{\left[2\left(t-2M_b^2\right)M_c^2-35\left(s-M_c^2\right)^2\right]\right.\nonumber\\
\al\al\left.\hspace{1cm}
+3\left[7\left(s-M_c^2\right)^2-\left(t-2M_b^2\right)M_c^2\right]\log\frac{M_c^2}{\mu^2}
\right\}\ .
\eea


\end{document}